\documentclass[11pt]{article}

\usepackage{palatino}
\usepackage{mathpazo}

\usepackage{braket} 
\usepackage{amsfonts}
\usepackage{amssymb}
\usepackage{amsmath}
\usepackage{latexsym}
\usepackage{amsthm}
\usepackage{eepic}
\usepackage{sectsty}
\usepackage[usenames]{color}
\usepackage{gastex} 
\usepackage{graphicx}

\newtheorem{theorem}{Theorem}

\newtheorem{prop}[theorem]{Proposition}
\newtheorem{cor}[theorem]{Corollary}
\theoremstyle{definition}
\newtheorem{definition}[theorem]{Definition}

\newcommand{\tinyspace}{\mspace{1mu}}
\newcommand{\op}[1]{\operatorname{#1}}
\newcommand{\abs}[1]{\left\lvert\tinyspace #1 \tinyspace\right\rvert}
\newcommand{\norm}[1]{\left\lVert\tinyspace#1\tinyspace\right\rVert}

\newcommand{\dnorm}[1]{\norm{#1}_{\diamond}}
\newcommand{\tr}{\operatorname{Tr}}
\newcommand{\ip}[2]{\left\langle #1 , #2\right\rangle}
\newcommand{\class}[1]{\textup{#1}}
\newcommand{\reg}[1]{\mathsf{#1}}

\def\real{\mathbb{R}}
\def\natural{\mathbb{N}}
\def\yes{\text{yes}}
\def\no{\text{no}}

\begin{document}
  
  \renewcommand \thesection{\Roman{section}}

  \begin{flushleft}
    {\huge \bf Quantum Computational Complexity}
  \end{flushleft}
  
  \vspace{2mm}
  
  \begin{flushleft} 
    {\Large John Watrous}\\[4mm]
    Institute for Quantum Computing and School of Computer Science\\
    University of Waterloo, Waterloo, Ontario, Canada.
  \end{flushleft}
  
  \section*{Article outline}
  
  \noindent
  \begin{tabular}{@{}ll@{}}
    {\bf I.}    & Definition of the subject and its importance\\[1mm]
    {\bf II.}   & Introduction\\[1mm]
    {\bf III.}  & The quantum circuit model\\[1mm]
    {\bf IV.}   & Polynomial-time quantum computations\\[1mm]
    {\bf V.}    & Quantum proofs\\[1mm]
    {\bf VI.}   & Quantum interactive proof systems\\[1mm]
    {\bf VII.}  & Other selected notions in quantum complexity\\[1mm]
    {\bf VIII.} & Future directions\\[1mm]
    {\bf IX.}   & References
  \end{tabular}
  
  \section*{Glossary}
  
  \begin{trivlist}
  \item {\bf Quantum circuit.} 
    \vspace{2mm}

    \noindent
    A quantum circuit is an acyclic network of quantum gates connected
    by wires: the gates represent quantum operations and the wires
    represent the qubits on which these operations are performed.
    The quantum circuit model is the most commonly studied model of
    quantum computation.
    \vspace{2mm}

  \item {\bf Quantum complexity class.}
    \vspace{2mm}

    \noindent
    A quantum complexity class is a collection of computational
    problems that are solvable by a chosen quantum computational model
    that obeys certain resource constraints.
    For example, \class{BQP} is the quantum complexity class of all
    decision problems that can be solved in polynomial time by a
    quantum computer.    
    \vspace{2mm}

  \item {\bf Quantum proof.}
    \vspace{2mm}

    \noindent
    A quantum proof is a quantum state that plays the role of a
    witness or certificate to a quantum computer that runs a
    verification procedure.
    The quantum complexity class \class{QMA} is defined by this
    notion: it includes all decision problems whose yes-instances are
    efficiently verifiable by means of quantum proofs.
    \vspace{2mm}
    
  \item {\bf Quantum interactive proof system.}
    \vspace{2mm}

    \noindent
    A quantum interactive proof system is an interaction between a
    verifier and one or more provers, involving the processing and
    exchange of quantum information, whereby the provers attempt to
    convince the verifier of the answer to some computational problem.
    
  \end{trivlist}
  
\section{Definition of the subject and its importance}

The inherent difficulty, or {\it hardness}, of computational problems
is a fundamental concept in computational complexity theory.
Hardness is typically formalized in terms of the resources required by
different models of computation to solve a given problem, such as the
number of steps of a deterministic Turing machine.
A variety of models and resources are often considered, including
deterministic, nondeterministic and probabilistic models; time and
space constraints; and interactions among models of differing
abilities.
Many interesting relationships among these different models and
resource constraints are known.

One common feature of the most commonly studied computational models
and resource constraint is that they are {\it physically motivated}.
This is quite natural, given that computers are physical devices, and
to a significant extent it is their study that motivates and directs
research on computational complexity.
The predominant example is the class of polynomial-time computable
functions, which ultimately derives its relevance from physical
considerations; for it is a mathematical abstraction of the class of
functions that can be efficiently computed without error by physical
computing devices.

In light of its close connection to the physical world, it seems only
natural that modern physical theories should be considered in the
context of computational complexity.
In particular, {\it quantum mechanics} is a clear candidate for a
physical theory to have the potential for implications, if not to
computational complexity then at least to computation more generally.
Given the steady decrease in the size of computing components, it
is inevitable that quantum mechanics will become increasingly relevant
to the construction of computers---for quantum mechanics provides a
remarkably accurate description of extremely small physical systems
(on the scale of atoms) where classical physical theories have failed
completely.
Indeed, an extrapolation of Moore's Law predicts sub-atomic computing
components within the next two decades \cite{Moore65,Lloyd00};
a possibility inconsistent with quantum mechanics as it is currently
understood.

That quantum mechanics should have implications to computational
complexity theory, however, is much less clear.
It is only through the remarkable discoveries and ideas of several
researchers, including Richard Feynman \cite{Feynman82}, David Deutsch
\cite{Deutsch85,Deutsch89}, Ethan Bernstein and Umesh Vazirani
\cite{BernsteinV93,BernsteinV97}, and Peter Shor \cite{Shor94,Shor97},
that this potential has become evident.
In particular, Shor's polynomial-time quantum factoring and
discrete-logarithm algorithms \cite{Shor97} give strong support to the
conjecture that quantum and classical computers yield differing
notions of computational hardness.
Other quantum complexity-theoretic concepts, such as the efficient
verification of quantum proofs, suggest a wider extent to which
quantum mechanics influences computational complexity.

It may be said that the principal aim of quantum computational
complexity theory is to understand the implications of quantum physics
to computational complexity theory.
To this end, it considers the hardness of computational problems with
respect to models of quantum computation, classifications of problems
based on these models, and their relationships to classical
models and complexity classes.

\section{Introduction}

This article surveys quantum computational complexity, with a focus on
three fundamental notions: polynomial-time quantum computations, the
efficient verification of quantum proofs, and quantum interactive
proof systems.
Based on these notions one defines quantum complexity classes,
such as \class{BQP}, \class{QMA}, and \class{QIP}, that contain
computational problems of varying hardness.
Properties of these complexity classes, and the relationships among
these classes and classical complexity classes, are presented.
As these notions and complexity classes are typically defined
within the quantum circuit model, this article includes a section that
focuses on basic properties of quantum circuits that are important in
the setting of quantum complexity.
A selection of other topics in quantum complexity, including quantum
advice, space-bounded quantum computation, and bounded-depth quantum
circuits, is also presented.

Two different but closely related areas of study are not discussed
in this article: {\it quantum query complexity} and
{\it quantum communication complexity}.
Readers interested in learning more about these interesting and active
areas of research may find the surveys of
Brassard \cite{Brassard03}, Cleve \cite{Cleve00a}, and de~Wolf
\cite{Wolf02} to be helpful starting points.

It is appropriate that brief discussions of computational complexity
theory and quantum information precede the main technical portion of
the article.
These discussions are intended only to highlight the aspects of these
topics that are non-standard, require clarification, or are of
particular importance in quantum computational complexity.
In the subsequent sections of this article, the reader is assumed to
have basic familiarity with both topics, which are covered in depth
by several text books
\cite{AroraB06,DuK00,KayeLM07,KitaevS+02,NielsenC00,Papadimitriou94}.

\subsection{Computational complexity}

\label{sec:complexity}

Throughout this article the binary alphabet $\{0,1\}$ is denoted
$\Sigma$, and all computational problems are assumed to be encoded
over this alphabet.
As usual, a function $f:\Sigma^{\ast}\rightarrow\Sigma^{\ast}$ is said
to be {\it polynomial-time computable} if there exists a polynomial-time
deterministic Turing machine that outputs $f(x)$ for every input
$x\in\Sigma^{\ast}$.
Two related points on the terminology used throughout this article are
as follows.
\begin{enumerate}
\item
A function of the form $p:\natural\rightarrow\natural$ (where
$\natural = \{0,1,2,\ldots\}$) is said to be a 
{\it polynomial-bounded function} if and only if there exists a
polynomial-time deterministic Turing machine that outputs $1^{f(n)}$
on input $1^n$ for every $n\in\natural$.
Such functions are upper-bounded by some polynomial, and are
efficiently computable.

\item
A function of the particular form $a:\natural\rightarrow [0,1]$ is
said to be {\it polynomial-time computable} if and only if there
exists a polynomial-time deterministic Turing machine that outputs a
binary representation of $a(n)$ on input $1^n$ for each
$n\in\natural$.
References to functions of this form in this article typically
concern bounds on probabilities that are functions of the length of an
input string to some problem.
\end{enumerate}

The notion of {\it promise problems} \cite{EvenS+84,Goldreich05} is
central to quantum computational complexity.
These are decision problems for which the input is assumed to be drawn
from some subset of all possible input strings.
More formally, a promise problem is a pair 
$A=(A_{\mathrm{yes}},A_{\mathrm{no}})$, where 
$A_{\mathrm{yes}},\,A_{\mathrm{no}}\subseteq\Sigma^{\ast}$ are sets
of strings satisfying $A_{\mathrm{yes}}\cap
A_{\mathrm{no}}=\varnothing$.
The strings contained in the sets $A_{\yes}$ and $A_{\no}$ are called
the {\it yes-instances} and {\it no-instances} of the problem, and
have answers {\it yes} and {\it no}, respectively.
{\it Languages} may be viewed as promise problems that obey the
additional constraint 
$A_{\mathrm{yes}}\cup A_{\mathrm{no}}=\Sigma^{\ast}$. 
Although complexity theory has traditionally focused on languages
rather than promise problems, little is lost and much is gained in
shifting one's focus to promise problems.
Karp reductions (also called polynomial-time many-to-one reductions)
and the notion of completeness are defined for promise problems in the
same way as for languages.

Several classical complexity classes are referred to in this article,
and compared with quantum complexity classes when relations are known.
The following classical complexity classes, which should hereafter be
understood to be classes of promise problems and not just languages,
are among those discussed.

\begin{center}
\begin{tabular}{p{0.6in}p{5.5in}}
{\bf \class{P}} &
A promise problem $A = (A_{\yes},A_{\no})$ is in \class{P} if and only
if there exists a polynomial-time deterministic Turing machine $M$
that accepts every string $x\in A_{\yes}$ and rejects every string
$x\in A_{\no}$.\\[3mm]
\end{tabular}
\begin{tabular}{p{0.6in}p{5.5in}}
{\bf \class{NP}} &
A promise problem $A = (A_{\yes},A_{\no})$ is in $\class{NP}$ if and
only if there exists a polynomial-bounded function $p$
and a polynomial-time deterministic Turing machine $M$ with the
following properties.
For every string $x\in A_{\yes}$, it holds that
$M$ accepts $(x,y)$ for some string $y\in\Sigma^{p(\abs{x})}$, and
for every string $x\in A_{\no}$, it holds that 
$M$ rejects $(x,y)$ for all strings $y\in\Sigma^{p(\abs{x})}$.
\\[3mm]
\end{tabular}
\begin{tabular}{p{0.6in}p{5.5in}}
{\bf \class{BPP}} & A promise problem $A = (A_{\yes},A_{\no})$ is in
\class{BPP} if and only if there exists a polynomial-time
probabilistic Turing machine $M$ that accepts every string $x\in
A_{\yes}$ with probability at least 2/3, and accepts every string
$x\in A_{\no}$ with probability at most 1/3.
\\[3mm]
\end{tabular}
\begin{tabular}{p{0.6in}p{5.5in}}
{\bf \class{PP}} & 
A promise problem $A = (A_{\yes},A_{\no})$ is in \class{PP} if and only
if there exists a polynomial-time probabilistic Turing machine $M$
that accepts every string $x\in A_{\yes}$ with probability strictly
greater than 1/2, and accepts every string $x\in A_{\no}$ with
probability at most 1/2.
\\[3mm]
\end{tabular}
\begin{tabular}{p{0.6in}p{5.5in}}
{\bf \class{MA}} & 
A promise problem $A = (A_{\yes},A_{\no})$ is in $\class{MA}$ if and
only if there exists a polynomial-bounded function $p$
and a probabilistic polynomial-time Turing machine $M$ with the
following properties.
For every string $x\in A_{\yes}$, it holds that
$\op{Pr}[\text{$M$ accepts $(x,y)$}] \geq \frac{2}{3}$
for some string $y\in\Sigma^{p(\abs{x})}$;
and for every string $x\in A_{\no}$, it holds that 
$\op{Pr}[\text{$M$ accepts $(x,y)$}] \leq \frac{1}{3}$
for all strings $y\in\Sigma^{p(\abs{x})}$.
\\[3mm]
\end{tabular}
\begin{tabular}{p{0.6in}p{5.5in}}
{\bf \class{AM}} & 
A promise problem $A = (A_{\yes},A_{\no})$ is in $\class{AM}$ if and
only if there exist polynomial-bounded functions $p$ and $q$ and a
polynomial-time deterministic Turing machine $M$ with the following 
properties.
For every string $x\in A_{\yes}$, and at least 2/3 of all strings
$y\in\Sigma^{p(\abs{x})}$, there exists a string
$z\in\Sigma^{q(\abs{x})}$ such that $M$ accepts $(x,y,z)$;
and for every string $x\in A_{\no}$, and at least 2/3 of all strings
$y\in\Sigma^{p(\abs{x})}$, there are no strings
$z\in\Sigma^{q(\abs{x})}$ such that $M$ accepts $(x,y,z)$.
\\[3mm]
\end{tabular}
\begin{tabular}{p{0.6in}p{5.5in}}
{\bf \class{SZK}} & A promise problem $A = (A_{\yes},A_{\no})$ is in
\class{SZK} if and only if it has a statistical zero-knowledge
interactive proof system.
\\[3mm]
\end{tabular}
\begin{tabular}{p{0.6in}p{5.5in}}
{\bf \class{PSPACE}} & A promise problem $A = (A_{\yes},A_{\no})$ is
in \class{PSPACE} if and only if there exists a deterministic Turing
machine $M$ running in polynomial space that accepts every string
$x\in A_{\yes}$ and rejects every string $x\in A_{\no}$.
\\[3mm]
\end{tabular}
\begin{tabular}{p{0.6in}p{5.5in}}
{\bf \class{EXP}} & A promise problem $A = (A_{\yes},A_{\no})$ is
in \class{EXP} if and only if there exists a deterministic Turing
machine $M$ running in exponential time (meaning time bounded by
$2^{p}$, for some polynomial-bounded function~$p$), that accepts every
string $x\in A_{\yes}$ and rejects every string $x\in A_{\no}$.
\\[3mm]
\end{tabular}
\begin{tabular}{p{0.6in}p{5.5in}}
  {\bf \class{NEXP}} &
  A promise problem $A = (A_{\yes},A_{\no})$ is in \class{NEXP} if and only
  if there exists an exponential-time non-deterministic Turing machine $N$
  for $A$.
  \\[3mm]
\end{tabular}
\begin{tabular}{p{0.6in}p{5.5in}}
  {\bf \class{PL}} & 
  A promise problem $A = (A_{\yes},A_{\no})$ is in \class{PL}
  if and only if there exists a probabilistic Turing machine $M$
  running in polynomial time and logarithmic space that accepts
  every string $x\in A_{\yes}$ with probability strictly greater
  than $1/2$ and accepts every string $x\in A_{\no}$ with
  probability at most $1/2$.
  \\[3mm] 
\end{tabular}
\begin{tabular}{p{0.6in}p{5.5in}}
  {\bf $\class{NC}$} & 
  A promise problem $A = (A_{\yes},A_{\no})$ is in $\class{NC}$
  if and only if there exists a logarithmic-space generated family
  $C = \{C_n\,:\,n\in\natural\}$ of poly-logarithmic depth Boolean
  circuits such that $C(x) = 1$ for all $x\in A_{\yes}$ and
  $C(x) = 0$ for all $x\in A_{\no}$.
\end{tabular}
\end{center}

For most of the complexity classes listed above, there is a standard
way to attach an {\it oracle} to the machine model that defines the
class, which provides a subroutine for solving instances
of a chosen problem $B = (B_{\yes},B_{\no})$.
One then denotes the existence of such an oracle with a
superscript---for example, $\class{P}^B$ is the class of promise
problems that can be solved in polynomial time by a deterministic
Turing machine equipped with an oracle that solves instances of $B$ (at
unit cost).
When classes of problems appear as superscripts, one takes the
union, as in the following example:
\[
\class{P}^{\class{NP}} = \bigcup_{B\in\class{NP}} \class{P}^B.
\]

\begin{figure}[t]
  \begin{center}
  \unitlength=1.4pt
  \begin{picture}(100, 215)(0,-50)
    \gasset{AHLength=0,ilength=-6}
    
    \node[Nframe=n](PL)(50,-50){\class{PL}}
    \node[Nframe=n](NC)(50,-25){\class{NC}}
    \node[Nframe=n](P)(50,0){\class{P}}
    \node[Nframe=n](NP)(75,25){\class{NP}}
    \node[Nframe=n](PP)(75,75){\class{PP}}
    \node[Nframe=n](BPP)(25,25){\class{BPP}}
    \node[Nframe=n](MA)(75,50){\class{MA}}
    \node[Nframe=n](SZK)(25,50){\class{SZK}}
    \node[Nframe=n](AM)(25,75){\class{AM}}
    \node[Nframe=n](PSPACE)(50,100){\class{PSPACE}}
    \node[Nframe=n](EXP)(50,125){\class{EXP}}
    \node[Nframe=n](NEXP)(50,150){\class{NEXP}}
    
    \drawedge(PL,NC){}
    \drawedge(NC,P){}
    \drawedge(P,BPP){}
    \drawedge(P,NP){}
    \drawedge(BPP,MA){}
    \drawedge(BPP,SZK){}
    \drawedge(SZK,AM){}
    \drawedge(NP,MA){}
    \drawedge(MA,AM){}
    \drawedge(MA,PP){}
    \drawedge(AM,PSPACE){}
    \drawedge(PP,PSPACE){}
    \drawedge(PSPACE,EXP){}
    \drawedge(EXP,NEXP){}
  \end{picture}
  \end{center}
  \caption{A diagram illustrating known inclusions among most of the
    classical complexity classes discussed in this paper.
    Lines indicate containments going upward; for example, \class{AM}
    is contained in \class{PSPACE}.}
  \label{fig:classical-classes}
\end{figure}
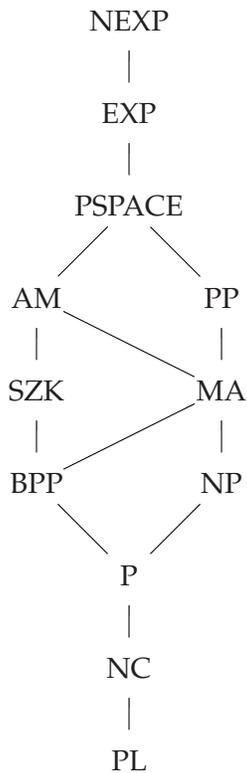

\subsection{Quantum information}

The standard general description of quantum information is used in
this article: mixed states of systems are represented by density
matrices
and operations are represented by completely positive trace-preserving
linear maps.
The choice to use this description of quantum information is deserving
of a brief discussion, for it will likely be less familiar to many
non-experts than the simplified picture of quantum information where
states are represented by unit vectors
and operations are represented by unitary matrices.
This simplified picture is indeed commonly used in the study of both
quantum algorithms and quantum complexity theory; and it is often
adequate.
However, the general picture has major advantages:
it unifies quantum information with classical probability theory,
brings with it powerful mathematical tools, and allows for simple and
intuitive descriptions in many situations where this is not possible
with the simplified picture.

Classical simulations of quantum computations, which are discussed
below in Section~\ref{sec:GapP}, may be better understood through a
fairly straightforward representation of quantum operations 
by matrices. 
This representation begins with a representation of density matrices
as vectors based on the function defined as 
$\op{vec}(\ket{x}\!\bra{y}) = \ket{x}\ket{y}$ for each choice
of $n\in\natural$ and $x,y\in\Sigma^n$, and extended by linearity to
all matrices indexed by $\Sigma^n$.
The effect of this mapping is to form a column vector by reading the
entries of a matrix in rows from left to right, starting at the top. 
For example,
\[
\op{vec}\begin{pmatrix}\alpha & \beta\\ \gamma & \delta \end{pmatrix}
=
\begin{pmatrix}\alpha\\ \beta\\ \gamma\\ \delta\end{pmatrix}.
\]
Now, the effect of a general quantum operation $\Phi$, represented in
the typical Kraus form as
\[
\Phi(\rho) = \sum_{j = 1}^k A_j \rho A_j^{\ast},
\]
is expressed as a matrix by means of the equality
\[
\op{vec}(\Phi(\rho)) 
= \left(\sum_{j = 1}^k A_j \otimes
\overline{A_j}\right)\op{vec}(\rho).
\]
The matrix
\[
K_{\Phi} = \sum_{j = 1}^k A_j \otimes\overline{A_j}
\]
is sometimes called the {\it natural representation} (or 
{\it linear representation}) of the operation $\Phi$.
Although this matrix could have negative or complex entries, one can
reasonably view it as being analogous to a stochastic matrix that
describes a probabilistic computation.

For example, the complete phase-damping channel for a single qubit can
be written
\[
D(\rho) = \ket{0}\braket{0|\rho|0}\bra{0}
+ \ket{1}\braket{1|\rho|1}\bra{1}.
\]
The effect of this mapping is to zero-out the off-diagonal entries of
a density matrix:
\[
D\begin{pmatrix}\alpha & \beta\\ \gamma & \delta \end{pmatrix}
=\begin{pmatrix}\alpha & 0\\ 0 & \delta \end{pmatrix}.
\]
The natural representation of this operation is easily computed:
\[
K_D
=
\begin{pmatrix}
  1&0&0&0\\[1mm]
  0&0&0&0\\[1mm]
  0&0&0&0\\[1mm]
  0&0&0&1
\end{pmatrix}.
\]

The natural matrix representation of quantum operations is well-suited
to performing computations.
For the purposes of this article, the most important observation about
this representation is that a composition of operations corresponds
simply to matrix multiplication.

\section{The quantum circuit model}

\subsection{General quantum circuits}

\begin{figure}[t]
  \begin{center}
    \unitlength=2.5pt
    \begin{picture}(140, 60)(0,0)
      \node[Nframe=n](In1)(-10,40){$\reg{X}_1$}
      \node[Nframe=n](In2)(-10,24){$\reg{X}_2$}
      \node[Nframe=n](In3)(-10,16){$\reg{X}_3$}
      \node[Nframe=n](In4)(-10,6){$\reg{X}_4$}
     
      \node[Nw=10,Nh=10,Nmr=0](1)(40,50){$\Phi_1$}
      \node[Nw=10,Nh=10,Nmr=0](2)(20,40){$\Phi_2$}
      \node[Nw=10,Nh=20,Nmr=0](3)(20,20){$\Phi_3$}
      \node[Nw=10,Nh=20,Nmr=0](4)(60,40){$\Phi_4$}
      \node[Nw=10,Nh=20,Nmr=0](5)(100,20){$\Phi_5$}
      \node[Nw=10,Nh=10,Nmr=0](6)(120,16){$\Phi_6$}
      
      \node[Nframe=n](Out1)(150,44){$\reg{Y}_1$}
      \node[Nframe=n](Out2)(150,24){$\reg{Y}_2$}
      \node[Nframe=n](Out3)(150,6){$\reg{Y}_3$}
      
      \drawedge(In1,2){}
      \drawedge[eyo=4](In2,3){}
      \drawedge[eyo=-4](In3,3){}
      \drawedge(In4,Out3){}
      
      \drawbpedge[syo=0,eyo=5](1,5,20,4,5,-20){}
      
      \drawedge(2,4){}
      
      \drawbpedge[syo=-5,eyo=-5](3,-6,40,4,-6,-40){} 
      \drawbpedge[eyo=-5,syo=5](3,5,30,5,5,-30){} 
      \drawedge(3,5){} 
      
      \drawbpedge[syo=-4,eyo=5](4,4,35,5,4,-35){} 
      \drawedge[syo=4](4,Out1){}
      
      \drawedge[syo=-4](5,6){}
      \drawedge[syo=4](5,Out2){}
    \end{picture}
  \end{center}
  \caption{An example of a quantum circuit.
    The input qubits are labelled $\reg{X}_1,\ldots,\reg{X}_4$, the
    output qubits are labelled $\reg{Y}_1,\ldots,\reg{Y}_3$, and the
    gates are labelled by (hypothetical) quantum operations 
    $\Phi_1,\ldots,\Phi_6$.}
  \label{fig:quantum-circuit}
\end{figure}
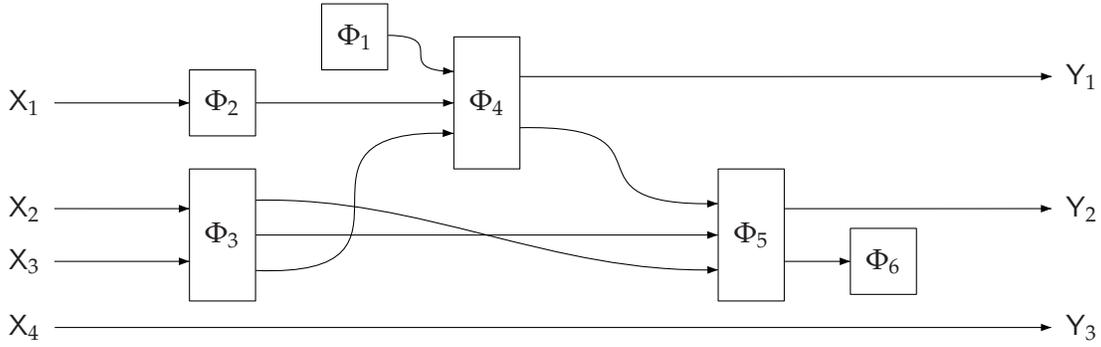

The term {\it quantum circuit} refers to an acyclic network of 
{\it quantum gates} connected by {\it wires}.
The quantum gates represent general quantum operations, involving some
constant number of qubits, while the wires represent the qubits on
which the gates act.
An example of a quantum circuit having four input qubits and three 
output qubits is pictured in Figure~\ref{fig:quantum-circuit}.
In general a quantum circuit may have $n$
input qubits and $m$ output qubits for any choice of integers $n,m\geq 0$.
Such a circuit induces some quantum operation from $n$ qubits to $m$
qubits, determined by composing the actions of the individual gates
in the appropriate way.
The {\it size} of a quantum circuit is the total number of gates plus
the total number of wires in the circuit.

A {\it unitary quantum circuit} is a quantum circuit in which all of
the gates correspond to unitary quantum operations.
Naturally this requires that every gate, and hence the circuit itself,
has an equal number of input and output qubits.
It is common in the study of quantum computing that one works
entirely with unitary quantum circuits.
The unitary model and general model are closely related, as will soon
be explained.

\subsection{A finite universal gate set} \label{sec:universal-gate-set}

Restrictions must be placed on the gates from which quantum circuits
may be composed if the quantum circuit model is to be used for
complexity theory---for without such restrictions it cannot be argued
that each quantum gate corresponds to an operation with unit-cost.
The usual way in which this is done is simply to fix a suitable finite
set of allowable gates.
For the remainder of this article, quantum circuits will be assumed to
be composed of gates from the following list:
\begin{enumerate}
\item[1.] {\it Toffoli gates}.
  Toffoli gates are three-qubit unitary gates defined by the following
  action on standard basis states:
  \[
  T : \ket{a}\ket{b}\ket{c} \mapsto \ket{a}\ket{b}\ket{c \oplus ab}.
  \]
  
\item[2.] {\it Hadamard gates}.
  Hadamard gates are single-qubit unitary gates defined by the
  following action on standard basis states:
  \[
  H : \ket{a} \mapsto \frac{1}{\sqrt{2}} \ket{0} + 
  \frac{(-1)^a}{\sqrt{2}} \ket{1}.
  \]
    
\item[3.] {\it Phase-shift gates}.
  Phase-shift gates are single-qubit unitary gates defined by the
  following action on standard basis states:
  \[
  P : \ket{a} \mapsto i^a \ket{a}.
  \]
  
\item[4.] {\it Ancillary gates}.
  Ancillary gates are non-unitary gates that take no input and
  produce a single qubit in the state $\ket{0}$ as output.
  
\item[5.] {\it Erasure gates}.
  Erasure gates are non-unitary gates that take a single qubit as
  input and produce no output.
  Their effect is represented by the partial trace on the space
  corresponding to the qubit they take as input.
  
\end{enumerate}

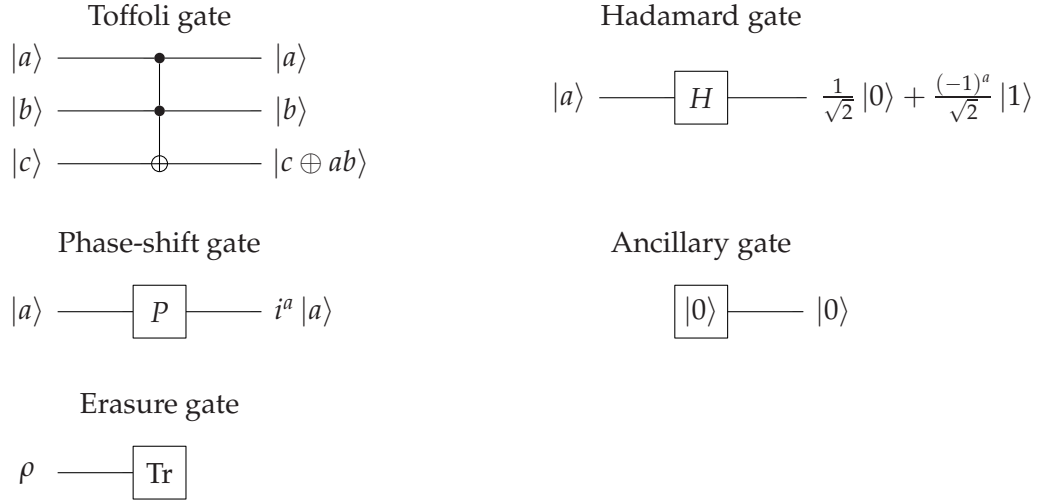
\begin{figure}[t]
  \begin{center}
    \begin{minipage}[t]{2.8in}
      \begin{center}
	\unitlength=1pt
	\begin{picture}(150, 70)(-20,15)
	  \gasset{AHLength=0,ilength=-6}
	  \put(50,75){\makebox(0,0){Toffoli gate}}
	  \node[Nframe=n,Nw=4,Nh=4,fillcolor=Black](T1)(50,60){} 
	  \node[Nframe=n,Nw=4,Nh=4,fillcolor=Black](T2)(50,40){} 
	  \node[Nw=6,Nh=6](T3)(50,20){} 
	  \imark[iangle= 90](T3)
	  \imark[iangle= 0](T3) 
	  \drawedge(T1,T2){}
	  \drawedge(T2,T3){}
          \node[Nframe=n](In1)(0,60){$\ket{a}$}
	  \node[Nframe=n](In2)(0,40){$\ket{b}$}
	  \node[Nframe=n](In3)(0,20){$\ket{c}$}
	  \node[Nframe=n](Out1)(100,60){$\ket{a}$}
	  \node[Nframe=n](Out2)(100,40){$\ket{b}$}
	  \node[Nframe=n,NLangle=0,NLdist=11](Out3)(100,20){%
	    $\ket{c\oplus ab}$}
	  \drawedge(In1,T1){}
	  \drawedge(In2,T2){}
	  \drawedge(In3,T3){}
	  \drawedge(T1,Out1){}
	  \drawedge(T2,Out2){}
	  \drawedge(T3,Out3){}
	\end{picture}
	\\
	\begin{picture}(150, 70)(-20,30)
	  \gasset{AHLength=0,ilength=-6}
	  \put(50,75){\makebox(0,0){Phase-shift gate}}
	  \node[Nw=20,Nh=20,Nmr=0](P)(50,50){$P$} 
	  \node[Nframe=n](In)(0,50){$\ket{a}$} 
	  \node[Nframe=n,NLangle=0,NLdist=5](Out)(100,50){%
	    $i^a\ket{a}$}
	  \drawedge(In,P){}
	  \drawedge(P,Out){}
	\end{picture}
	\\
	\begin{picture}(150, 50)(-20,40)
	  \gasset{AHLength=0,ilength=-6}
	  \put(50,75){\makebox(0,0){Erasure gate}}
	  \node[Nw=20,Nh=20,Nmr=0](Tr)(50,50){$\tr$} 
	  \node[Nframe=n](In)(0,50){$\rho$}
	  \drawedge(In,Tr){}
	\end{picture}
      \end{center}
    \end{minipage}
    \hspace*{6mm}
    \begin{minipage}[t]{2.8in}
      \begin{center}
	\unitlength=1pt
	\begin{picture}(150, 70)(0,15)
	  \gasset{AHLength=0,ilength=-6}
	  \put(50,75){\makebox(0,0){Hadamard gate}}
	  \node[Nw=20,Nh=20,Nmr=0](H)(50,45){$H$} 
	  \node[Nframe=n](In)(0,45){$\ket{a}$} 
	  \node[Nframe=n,Nw=10](Out)(95,45){%
	    \makebox(0,0)[l]{$\frac{1}{\sqrt{2}}\ket{0} +
	      \frac{(-1)^a}{\sqrt{2}}\ket{1}$}}
	  \drawedge(In,H){}
	  \drawedge(H,Out){}
	\end{picture}
	\\
	\begin{picture}(150, 70)(0,30)
	  \gasset{AHLength=0,ilength=-6}
	  \put(50,75){\makebox(0,0){Ancillary gate}}
	  \node[Nw=20,Nh=20,Nmr=0](0)(50,50){$\ket{0}$} 
	  \node[Nframe=n](Out)(100,50){$\ket{0}$}
	  \drawedge(0,Out){}
	\end{picture}
      \end{center}
    \end{minipage}
  \end{center}

  \caption{A universal collection of quantum gates: 
    Toffoli, Hadamard, phase-shift, ancillary, and erasure gates.}
  \label{fig:gates}
\end{figure}

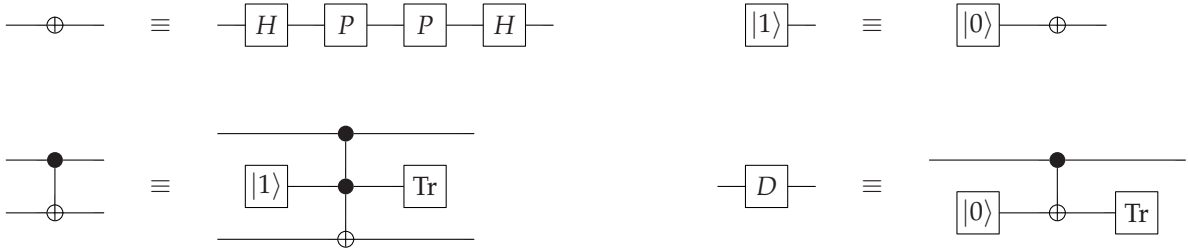
\begin{figure}[t]
  \begin{center}
    \unitlength=1pt
    \small
    \begin{picture}(220, 50)(10,0)
      \gasset{AHLength=0,ilength=-6}
      
      \node[Nframe=n](In1)(0,20){}
      \node[Nw=6,Nh=6](1)(30,20){} 
      \imark[iangle= 90](1)
      \imark[iangle= 0](1) 

      \node[Nframe=n](Out1)(60,20){}
      \drawedge(In1,1){}
      \drawedge(1,Out1){}
      
      \node[Nframe=n](Symbol)(70,20){$\equiv$}
      
      \node[Nframe=n](In2)(80,20){}
      \node[Nw=16,Nh=16,Nmr=0](H1)(110,20){$H$}
      \node[Nw=16,Nh=16,Nmr=0](P1)(140,20){$P$}
      \node[Nw=16,Nh=16,Nmr=0](P2)(170,20){$P$}
      \node[Nw=16,Nh=16,Nmr=0](H2)(200,20){$H$}
      \node[Nframe=n](Out2)(230,20){}
      
      \drawedge(In2,H1){}
      \drawedge(H1,P1){}
      \drawedge(P1,P2){}
      \drawedge(P2,H2){}
      \drawedge(H2,Out2){}
    \end{picture}
    \hspace{5mm}
    \begin{picture}(220, 50)(-20,0)
      \gasset{AHLength=0,ilength=-6}
      
      \node[Nw=16,Nh=16,Nmr=0](1)(30,20){$\ket{1}$}
      \node[Nframe=n](Out1)(60,20){}
      \drawedge(1,Out1){}
      
      \node[Nframe=n](Symbol)(70,20){$\equiv$}
      
      \node[Nw=16,Nh=16,Nmr=0](0)(110,20){$\ket{0}$}

      \node[Nw=6,Nh=6](X)(140,20){} 
      \imark[iangle= 90](X)
      \imark[iangle= 0](X) 

      \node[Nframe=n](Out2)(170,20){}
      
      \drawedge(0,X){}
      \drawedge(X,Out2){}
      
    \end{picture}

    \begin{picture}(220, 60)(10,20)
      
      \gasset{AHLength=0,ilength=-6}
      
      \node[Nframe=n,Nw=6,Nh=6,fillcolor=Black](CN1)(30,50){} 
      \node[Nw=6,Nh=6](CN2)(30,30){} 
      \imark[iangle= 90](CN2)
      \imark[iangle= 0](CN2) 
      \drawedge(CN1,CN2){}
      \node[Nframe=n](In4)(0,50){}	
      \node[Nframe=n](In5)(0,30){}	
      \node[Nframe=n](Out4)(60,50){}	
      \node[Nframe=n](Out5)(60,30){}	
      \drawedge(In4,CN1){}
      \drawedge(In5,CN2){}
      \drawedge(CN1,Out4){}
      \drawedge(CN2,Out5){}
      
      \node[Nframe=n](Symbol)(70,40){$\equiv$}
      
      \node[Nframe=n](In1)(80,60){}
      \node[Nframe=n](In3)(80,20){}
      
      \node[Nframe=n](Out1)(200,60){}	
      \node[Nframe=n](Out3)(200,20){}	
      
      \node[Nframe=n,Nw=6,Nh=6,fillcolor=Black](T1)(140,60){} 
      \node[Nframe=n,Nw=6,Nh=6,fillcolor=Black](T2)(140,40){} 
      \node[Nw=6,Nh=6](T3)(140,20){} 
      \imark[iangle= 90](T3)
      \imark[iangle= 0](T3) 
      \drawedge(T1,T2){}
      \drawedge(T2,T3){}
      
      \node[Nw=16,Nh=16,Nmr=0](1)(110,40){$\ket{1}$}
      \node[Nw=16,Nh=16,Nmr=0](Tr2)(170,40){$\tr$}
      
      \drawedge(In1,T1){}
      \drawedge(In3,T3){}
      \drawedge(1,T2){}
      \drawedge(T3,Out3){}
      \drawedge(T1,Out1){}
      \drawedge(T2,Tr2){}
         
    \end{picture}
    \hspace{5mm}
    \begin{picture}(220, 60)(-110,20)
      \gasset{AHLength=0,ilength=-6}
      
      \node[Nframe=n](In)(-90,40){}
      \node[Nframe=n](Out)(-30,40){}
      \node[Nw=16,Nh=16,Nmr=0](D)(-60,40){$D$}
      \drawedge(In,D){}
      \drawedge(D,Out){}
      
      \node[Nframe=n](Symbol)(-20,40){$\equiv$}
      
      \node[Nframe=n,Nw=6,Nh=6,fillcolor=Black](T1)(50,50){} 
      \node[Nw=6,Nh=6](T2)(50,30){} 
      \imark[iangle= 90](T2)
      \imark[iangle= 0](T2) 
      \drawedge(T1,T2){}
      
      \node[Nframe=n](In1)(-10,50){}
      \node[Nw=16,Nh=16,Nmr=0](0)(20,30){$\ket{0}$}
      \node[Nw=16,Nh=16,Nmr=0](Tr2)(80,30){$\tr$}
      \node[Nframe=n](Out1)(110,50){}
      
      \drawedge(In1,T1){}
      \drawedge(0,T2){}
      \drawedge(T2,Tr2){}
      \drawedge(T1,Out1){}
      
    \end{picture}
  \end{center}
  \caption{Four additional quantum gates, together with their
    implementations as quantum circuits.
    Top left: a NOT gate.
    Top right: a constant $\ket{1}$ ancillary gate.
    Bottom left: a controlled-NOT gate.
    Bottom right: a phase-damping (or decoherence) gate.}
  \label{fig:additional-gates}
\end{figure}

\noindent
The symbols used to denote these gates in quantum circuit diagrams are
shown in Figure~\ref{fig:gates}.
Some additional useful gates are illustrated in
Figure~\ref{fig:additional-gates}, along with their realizations as
circuits with gates from the chosen basis set.

The above gate set is {\it universal} in a strong sense:
every quantum operation can be approximated to within any
desired degree of accuracy by some quantum circuit.
Moreover, the size of the approximating circuit scales well with
respect to the desired accuracy.
Theorem~\ref{theorem:universality}, stated below, expresses this fact
in more precise terms, but requires a brief discussion of a specific
sense in which one quantum operation approximates another.

A natural and operationally meaningful way to measure the distance
between two given quantum operations $\Phi$ and $\Psi$ is given by
\[
\delta(\Phi,\Psi) = \frac{1}{2}\dnorm{\Phi-\Psi},
\]
where $\dnorm{\cdot}$ denotes a norm usually known as the
{\it diamond norm} \cite{Kitaev97,KitaevS+02}.
A technical discussion of this norm is not necessary for the purposes
of this article and is beyond its scope.
Instead, an intuitive description of the distance measure
$\delta(\Phi,\Psi)$ will suffice.

When considering the distance between quantum operations $\Phi$ and
$\Psi$, it must naturally be assumed that these operations agree on their
numbers of input qubits and output qubits;
so assume that $\Phi$ and $\Psi$ both map $n$ qubits to $m$ qubits for
nonnegative integers $n$ and $m$.
Now, suppose that an arbitrary quantum state on $n$ or more qubits is
prepared, one of the two operations $\Phi$ or $\Psi$ is applied to the
first $n$ of these qubits, and then a general measurement of all of
the resulting qubits takes place (including the $m$ output qubits and the
qubits that were not among the inputs to the chosen quantum
operation).
Two possible probability distributions on measurement outcomes arise:
one corresponding to $\Phi$ and the other corresponding to $\Psi$.
The quantity $\delta(\Phi,\Psi)$ is simply the maximum possible total
variation distance between these distributions, ranging over all
possible initial states and general measurements.
This is a number between 0 and 1 that intuitively represents the
{\it observable difference} between quantum operations.
In the special case that $\Phi$ and $\Psi$ take no inputs, the
quantity $\delta(\Phi,\Psi)$ is simply one-half the trace norm of the
difference between the two output states; a common and useful ways to
measure the distance between quantum states.

Now the Universality Theorem, which represents an amalgamation of
several results that suits the needs of this article, may be stated.
In particular, it incorporates the Solovay--Kitaev Theorem, which
provides a bound on the size of an approximating circuit as a
function of the accuracy.

\begin{theorem}[Universality Theorem] \label{theorem:universality}
  Let $\Phi$ be an arbitrary quantum operation from $n$ qubits to $m$
  qubits. 
  Then for every $\varepsilon>0$ there exists a quantum circuit $Q$
  with $n$ input qubits and $m$ output qubits such that 
  $\delta(\Phi,Q) < \varepsilon$.
  Moreover, for fixed $n$ and $m$, the circuit $Q$ may be taken to
  satisfy
  $\op{size}(Q) = \op{poly}(\log(1/\varepsilon))$.
\end{theorem}
    
\noindent
Note that it is inevitable that the size of $Q$ is exponential in $n$
and $m$ in the worst case \cite{Knill95a}.
Further details on the facts comprising this theorem can be found in
Nielsen and Chuang \cite{NielsenC00} and Kitaev, Shen, and Vyalyi
\cite{KitaevS+02}.

\subsection{Unitary purifications of quantum circuits}

The connection between the general and unitary quantum circuits
can be understood through the notion of a 
{\it unitary purification} of a general quantum circuit. 
This may be thought of as a very specific manifestation of the 
{\it Stinespring Dilation Theorem} \cite{Stinespring55}, which implies
that general quantum operations can be represented by unitary operations
on larger systems.
It was first applied to the quantum circuit model
by Aharonov, Kitaev, and Nisan \cite{AharonovK+98}, who gave several
arguments in favor of the general quantum circuit model over the
unitary model.
The term {\it purification} is borrowed from the notion of a
purification of a mixed quantum state, as the process of
unitary purification for circuits is similar in spirit.
The universal gate described in the previous section has the effect of
making the notion of a unitary purification of a general quantum
circuit nearly trivial at a technical level.

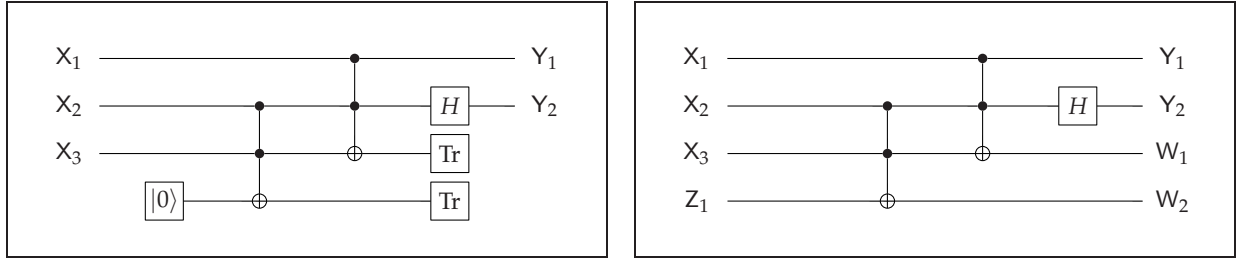
\begin{figure}[t]
  \begin{center}
    \unitlength=0.9pt
    \footnotesize
    \fbox{
      \begin{picture}(240, 100)(-20,-20)
	\gasset{AHLength=0,ilength=-6}
	
	\node[Nframe=n](In1)(0,60){$\reg{X}_1$}
	\node[Nframe=n](In2)(0,40){$\reg{X}_2$}
	\node[Nframe=n](In3)(0,20){$\reg{X}_3$}

	\node[Nw=16,Nh=16,Nmr=0](0)(40,0){$\ket{0}$}
	
	\node[Nframe=n,Nw=4,Nh=4,fillcolor=Black](T11)(80,40){} 
	\node[Nframe=n,Nw=4,Nh=4,fillcolor=Black](T12)(80,20){} 
	\node[Nw=6,Nh=6](T13)(80,0){} 
	\imark[iangle= 90](T13)
	\imark[iangle= 0](T13) 
	\drawedge(T11,T12){}
	\drawedge(T12,T13){}

	\node[Nframe=n,Nw=4,Nh=4,fillcolor=Black](T21)(120,60){} 
	\node[Nframe=n,Nw=4,Nh=4,fillcolor=Black](T22)(120,40){} 
	\node[Nw=6,Nh=6](T23)(120,20){} 
	\imark[iangle= 90](T23)
	\imark[iangle= 0](T23) 
	\drawedge(T21,T22){}
	\drawedge(T22,T23){}
        
	\node[Nw=16,Nh=16,Nmr=0](H)(160,40){$H$}
	\node[Nw=16,Nh=16,Nmr=0](Tr1)(160,20){$\tr$}
	\node[Nw=16,Nh=16,Nmr=0](Tr2)(160,0){$\tr$}
	\node[Nframe=n](Out1)(200,60){$\reg{Y}_1$}
	\node[Nframe=n](Out2)(200,40){$\reg{Y}_2$}

	\drawedge(In1,T21){}
	\drawedge(In2,T11){}
	\drawedge(In3,T12){}
	\drawedge(0,T13){}
	\drawedge(T11,T22){}
	\drawedge(T12,T23){}
	\drawedge(T13,Tr2){}
	\drawedge(T21,Out1){}
	\drawedge(T22,H){}
	\drawedge(T23,Tr1){}
	\drawedge(H,Out2){}
      \end{picture}
    }
    \rule{2mm}{0mm}
    \fbox{
      \begin{picture}(240, 100)(-20,-20)
	\gasset{AHLength=0,ilength=-6}
	
	\node[Nframe=n](In1)(0,60){$\reg{X}_1$}
	\node[Nframe=n](In2)(0,40){$\reg{X}_2$}
	\node[Nframe=n](In3)(0,20){$\reg{X}_3$}

	\node[Nframe=n,Nw=4,Nh=4,fillcolor=Black](T11)(80,40){} 
	\node[Nframe=n,Nw=4,Nh=4,fillcolor=Black](T12)(80,20){} 
	\node[Nw=6,Nh=6](T13)(80,0){} 
	\imark[iangle= 90](T13)
	\imark[iangle= 0](T13) 
	\drawedge(T11,T12){}
	\drawedge(T12,T13){}

	\node[Nframe=n,Nw=4,Nh=4,fillcolor=Black](T21)(120,60){} 
	\node[Nframe=n,Nw=4,Nh=4,fillcolor=Black](T22)(120,40){} 
	\node[Nw=6,Nh=6](T23)(120,20){} 
	\imark[iangle= 90](T23)
	\imark[iangle= 0](T23) 
	\drawedge(T21,T22){}
	\drawedge(T22,T23){}
        
	\node[Nw=16,Nh=16,Nmr=0](H)(160,40){$H$}
	\node[Nframe=n](Out1)(200,60){$\reg{Y}_1$}
	\node[Nframe=n](Out2)(200,40){$\reg{Y}_2$}

	\node[Nframe=n](In4)(0,0){$\reg{Z}_1$}
	\node[Nframe=n](Out3)(200,20){$\reg{W}_1$}
	\node[Nframe=n](Out4)(200,0){$\reg{W}_2$}

	\drawedge(In1,T21){}
	\drawedge(In2,T11){}
	\drawedge(In3,T12){}
	\drawedge(In4,T13){}
	\drawedge(T11,T22){}
	\drawedge(T12,T23){}
	\drawedge(T13,Out4){}
	\drawedge(T21,Out1){}
	\drawedge(T22,H){}
	\drawedge(T23,Out3){}
	\drawedge(H,Out2){}
	
      \end{picture} 
    }
  \end{center}

  \caption{A general quantum circuit (left) and its unitary
    purification (right).}
  \label{fig:purification}
\end{figure}

Suppose that $Q$ is a quantum circuit taking input qubits
$(\reg{X}_1,\ldots,\reg{X}_n)$ and producing output qubits
$(\reg{Y}_1,\ldots,\reg{Y}_m)$, and assume there are $k$ ancillary
gates and $l$ erasure gates among the gates of $Q$ to be labelled in
an arbitrary order as $G_1,\ldots,G_k$ and $K_1,\ldots,K_l$,
respectively. 
A new quantum circuit $R$ may then be formed by removing the gates
labelled $G_1,\ldots,G_k$ and $K_1,\ldots,K_l$; and to account for the
removal of these gates the circuit $R$ takes $k$ additional input
qubits $(\reg{Z}_1,\ldots,\reg{Z}_k)$ and produces $l$ additional
output qubits $(\reg{W}_1,\ldots,\reg{W}_l)$.
Figure~\ref{fig:purification} illustrates this process.
The circuit $R$ is said to be a unitary purification of $Q$.
It is obvious that $R$ is equivalent to $Q$, provided the qubits 
$(\reg{Z}_1,\ldots,\reg{Z}_k)$ are initially set to the $\ket{0}$
state and the qubits $(\reg{W}_1,\ldots,\reg{W}_l)$ are traced-out, or
simply ignored, after the circuit is run---for this is precisely the
meaning of the removed gates.

Despite the simplicity of this process, it is often useful to consider
the properties of unitary purifications of general quantum circuits.

\subsection{Oracles in the quantum circuit model}

\label{sec:oracle-gates}

Oracles play an important, and yet uncertain, role in computational
complexity theory; and the situation is no different in the quantum
setting.
Several interesting oracle-related results, offering some insight into
the power of quantum computation, will be discussed in this article.

Oracle queries are represented in the quantum circuit model
by an infinite family 
\[
\{R_n\,:\,n\in\natural\}
\]
of quantum gates, one for each possible query length.
Each gate $R_n$ is a unitary gate acting on $n+1$ qubits, with the 
effect on computational basis states given by
\begin{equation} \label{eq:query}
R_n \ket{x,a} = \ket{x, a\oplus A(x)}
\end{equation}
for all $x\in\Sigma^n$ and $a\in\Sigma$, where $A$ is some predicate
that represents the particular oracle under consideration.
When quantum computations relative to such an oracle are to be
studied, quantum circuits composed of ordinary quantum gates as well
as the oracle gates $\{R_n\}$ are considered; the interpretation being
that each instance of $R_n$ in such a circuit represents one oracle
query.

It is critical to many results concerning quantum oracles, as well as
most results in the area of quantum query complexity, that the above
definition \eqref{eq:query} takes each $R_n$ to be unitary, thereby
allowing these gates to make queries ``in superposition''.
In support of this seemingly strong definition is the fact, discussed
in the next section, that any efficient algorithm (quantum or
classical) can be converted to a quantum circuit that closely
approximates the action of these gates.

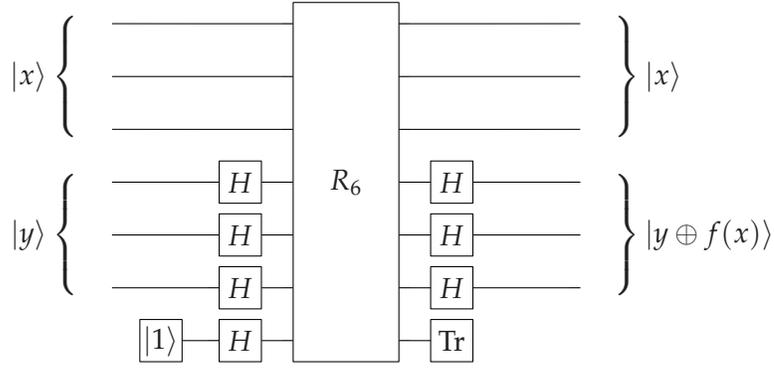
\begin{figure}[t]
  \begin{center}
    \unitlength=1pt
    \begin{picture}(100, 150)(0,30)
      \gasset{AHLength=0,ilength=-6}
      
      \node[Nw=40,Nh=136,Nmr=0](R)(50,100){$R_6$}
      \node[Nw=16,Nh=16,Nmr=0](H1)(10,100){$H$}
      \node[Nw=16,Nh=16,Nmr=0](H2)(10,80){$H$}
      \node[Nw=16,Nh=16,Nmr=0](H3)(10,60){$H$}
      \node[Nw=16,Nh=16,Nmr=0](H4)(10,40){$H$}
      \node[Nw=16,Nh=16,Nmr=0](H5)(90,100){$H$}
      \node[Nw=16,Nh=16,Nmr=0](H6)(90,80){$H$}
      \node[Nw=16,Nh=16,Nmr=0](H7)(90,60){$H$}
      \node[Nw=16,Nh=16,Nmr=0](Tr)(90,40){$\tr$}
      \node[Nw=16,Nh=16,Nmr=0](1)(-20,40){$\ket{1}$}
      \node[Nframe=n](In1)(-50,160){}
      \node[Nframe=n](In2)(-50,140){}
      \node[Nframe=n](In3)(-50,120){}
      \node[Nframe=n](In4)(-50,100){}
      \node[Nframe=n](In5)(-50,80){}
      \node[Nframe=n](In6)(-50,60){}
      \node[Nframe=n](In7)(-50,40){}
      \node[Nframe=n](Out1)(150,160){}
      \node[Nframe=n](Out2)(150,140){}
      \node[Nframe=n](Out3)(150,120){}
      \node[Nframe=n](Out4)(150,100){}
      \node[Nframe=n](Out5)(150,80){}
      \node[Nframe=n](Out6)(150,60){}
      \node[Nframe=n](Out7)(150,40){}
      \drawedge[eyo=60](In1,R){}
      \drawedge[eyo=40](In2,R){}
      \drawedge[eyo=20](In3,R){}
      \drawedge[eyo=0](H1,R){}
      \drawedge[eyo=-20](H2,R){}
      \drawedge[eyo=-40](H3,R){}
      \drawedge[eyo=-60](H4,R){}
      \drawedge[eyo=60](Out1,R){}
      \drawedge[eyo=40](Out2,R){}
      \drawedge[eyo=20](Out3,R){}
      \drawedge(In4,H1){}
      \drawedge(In5,H2){}
      \drawedge(In6,H3){}
      \drawedge(H5,Out4){}
      \drawedge(H6,Out5){}
      \drawedge(H7,Out6){}
      \drawedge(1,H4){}
      \drawedge[eyo=0](H5,R){}
      \drawedge[eyo=-20](H6,R){}
      \drawedge[eyo=-40](H7,R){}
      \drawedge[eyo=-60](Tr,R){}
      \put(-50,140){\makebox(0,0)[r]{$\ket{x}\left\{\rule{0mm}{9mm}\right.$}}
      \put(-50,80){\makebox(0,0)[r]{$\ket{y}\left\{\rule{0mm}{9mm}\right.$}}
      \put(150,140){\makebox(0,0)[l]{$\left.\rule{0mm}{9mm}\right\}\ket{x}$}}
      \put(150,80){\makebox(0,0)[l]{$\left.\rule{0mm}{9mm}\right\}
	  \ket{y \oplus f(x)}$}}
    \end{picture}
  \end{center}
  \caption{The Bernstein--Vazirani algorithm allows a multiple-bit
    query to be simulated by a single-bit query.
    In the example pictured, $f:\Sigma^3\rightarrow\Sigma^3$ is a
    given function.
    To simulate a query to this function, the gate $R_6$ is taken to
    be a standard oracle gate implementing the predicate
    $A(x,z) = \ip{f(x)}{z}$, for $\ip{\cdot}{\cdot}$ denoting the
    modulo 2 inner product.}
  \label{fig:BV}
\end{figure}

Finally, one may consider a more general situation in which the
predicate $A$ is replaced by a function that outputs multiple bits.
The definition of each gate $R_n$ is adapted appropriately.
Alternately, one may restrict their attention to single-bit queries as
discussed above, and use the Bernstein--Vazirani algorithm
\cite{BernsteinV97} to simulate one multiple-bit query with one
single-bit query as illustrated in Figure~\ref{fig:BV}.

\section{Polynomial-time quantum computations}

This section focuses on {\it polynomial-time quantum computations}.
These are the computations that are viewed, in an abstract and idealized
sense, to be efficiently implementable by the means of a quantum
computer.
In particular, the complexity class \class{BQP} (short for 
{\it bounded-error quantum polynomial time}) is defined.
This is the most fundamentally important of all quantum complexity
classes, as it represents the collection of decision problems that can be
efficiently solved by quantum computers.

\subsection{Polynomial-time generated circuit families and
  \class{BQP}}

To define the class \class{BQP} using the quantum circuit model, it is
necessary to briefly discuss encodings of circuits and the notion of a
polynomial-time generated circuit family.

It is clear that any quantum circuit formed from the gates described
in the previous section could be {\it encoded} as a binary string
using any number of different encoding schemes.
Such an encoding scheme must be chosen, but its specifics are not
important so long as the following simple restrictions are satisfied:
\begin{enumerate}
\item[1.]
  The encoding is sensible: every quantum circuit is encoded by at
  least one binary string, and every binary string encodes at most one
  quantum circuit.  

\item[2.]
  The encoding is efficient:
  there is a fixed polynomial-bounded function $p$ such that every
  circuit of size $N$ has an encoding with length at most $p(N)$.
  Specific information about the structure of a circuit must be
  computable in polynomial time from an encoding of the circuit.

\item[3.]
  The encoding disallows compression: it is not possible to work with
  encoding schemes that allow for extremely short (e.g.,
  polylogarithmic-length) encodings of circuits;
  so for simplicity it is assumed that the length of every encoding of
  a quantum circuit is at least the size of the circuit.
\end{enumerate}

Now, as any quantum circuit represents a finite computation with some
fixed number of input and output qubits, quantum algorithms are
modelled by {\it families} of quantum circuits.
The typical assumption is that a quantum circuit family that describes
an algorithm contains one circuit for each possible input length.
Precisely the same situation arises here as in the classical setting,
which is that it should be possible to efficiently generate the
circuits in a given family in order for that family to represent
an efficient, finitely specified algorithm.
The following definition formalizes this notion.

\begin{definition}
  Let $S\subseteq\Sigma^{\ast}$ be any set of strings.
  Then a collection $\left\{ Q_x\,:\,x\in S\right\}$
  of quantum circuits is said to be {\it polynomial-time generated}
  if there exists a polynomial-time deterministic Turing machine
  that, on every input $x\in S$, outputs an encoding of $Q_x$.
\end{definition}

This definition is slightly more general than what is needed to define
\class{BQP}, but is convenient for other purposes.
For instance, it allows one to easily consider the situation in which
the input, or some part of the input, for some problem is hard-coded
into a collection of circuits; or where a computation for some input
may be divided among several circuits.
In the most typical case that a polynomial-time generated family of
the form $\{Q_n\,:\,n\in\natural\}$ is referred to, it should be
interpreted that this is a shorthand for $\{Q_{1^n}\,:\,n\in\natural\}$.
Notice that every polynomial-time generated family 
$\{Q_x\,:\,x\in S\}$ has the property that each circuit $Q_x$ has size
polynomial in $\abs{x}$.
Intuitively speaking, the number of quantum and classical computation
steps required to implement such a computation is polynomial; and so
operations induced by the circuits in such a family are viewed as 
representing {\it polynomial-time quantum computations}.

  

The complexity class \class{BQP}, which contains those promise
problems abstractly viewed to be efficiently solvable using a quantum
computer, may now be defined.
More precisely, \class{BQP} is the class of promise problems that can
be solved by polynomial-time quantum computations that may have some
small probability to make an error.
For decision problems, the notion of a polynomial-time quantum
computation is equated with the computation of a polynomial-time
generated quantum circuit family $Q = \{Q_n\,:\,n\in\natural\}$, where
each circuit $Q_n$ takes $n$ input qubits, and produces one output
qubit.
The computation on a given input string $x\in\Sigma^{\ast}$ is
obtained by first applying the circuit $Q_{\abs{x}}$ to the state
$\ket{x}\!\!\bra{x}$, and then measuring the output qubit with respect
to the standard basis.
The measurement results 0 and 1 are interpreted as {\it yes} and 
{\it no} (or {\it accept} and {\it reject}), respectively.
The events that $Q$ accepts $x$ and $Q$ rejects $x$ are understood to
have associated probabilities determined in this way.

\begin{center}
\begin{tabular}{p{0.4in}p{5.7in}}
{\bf \class{BQP}} & 
Let $A = (A_{\yes},A_{\no})$ be a promise problem and let
$a,b:\natural \rightarrow[0,1]$ be functions.
Then $A\in\class{BQP}(a,b)$ if and only if there
exists a polynomial-time generated family of quantum circuits
$Q = \{Q_n\,:\,n\in\natural\}$, where each circuit $Q_n$ takes $n$
input qubits and produces one output qubit, that satisfies the
following properties:
\begin{enumerate}
\item if $x\in A_{\yes}$ then 
    $\op{Pr}[\text{$Q$ accepts $x$}]\geq a(\abs{x})$, and 
\item if $x\in A_{\no}$ then 
  $\op{Pr}[\text{$Q$ accepts $x$}]\leq b(\abs{x})$.
\end{enumerate}
The class \class{BQP} is defined as $\class{BQP} = \class{BQP}(2/3,1/3)$.
\end{tabular}
\end{center}

\noindent
Similar to \class{BPP}, there is nothing special about the particular
choice of error probability 1/3, other than that it is a constant
strictly smaller than 1/2.
This is made clear in the next section.

There are several problems known to be in \class{BQP} but not known
(and generally not believed) to be in \class{BPP}.
Decision-problem variants of the integer factoring and discrete
logarithm problems, shown to be in \class{BQP} by Shor \cite{Shor97},
are at present the most important and well-known examples.

\subsection{Error reduction for \class{BQP}}

When one speaks of the flexibility, or {\it robustness}, of
\class{BQP} with respect to error bounds, it is meant that the class
$\class{BQP}(a,b)$ is invariant under a wide range of ``reasonable''
choices of the functions $a$ and $b$.
The following proposition states this more precisely.  

\begin{prop}[Error reduction for \class{BQP}]
  \label{prop:BQP-error}
  Suppose that $a,b:\natural \rightarrow[0,1]$ are polynomial-time
  computable functions and $p:\natural\rightarrow\natural$ is a
  polynomial-bounded function such that
  $a(n) - b(n) \geq 1/p(n)$ for all but finitely many $n\in\natural$.
  Then for every choice of a polynomial-bounded function
  $q:\natural\rightarrow\natural$ satisfying $q(n) \geq 2$ for all
  but finitely many $n\in\natural$, it holds that
  \[
  \class{BQP}(a,b) = \class{BQP} =
  \class{BQP}\left(1-2^{-q},2^{-q}\right).
  \]
\end{prop}

The above proposition may be proved in the same standard way that
similar statements are proved for classical probabilistic
computations: by repeating a given computation some large (but still
polynomial) number of times, overwhelming statistical evidence is
obtained so as to give the correct answer with an extremely small
probability of error.
It is straightforward to represent this sort of repeated computation
within the quantum circuit model in such a way that the requirements
of the definition of \class{BQP} are satisfied.

\subsection{Simulating classical computations with quantum circuits}

It should not be surprising that quantum computers can
efficiently simulate classical computers---for quantum information
generalizes classical information, and it would be absurd if there
were a loss of computational power in moving to a more general model.
This intuition may be confirmed by observing the containment
$\class{BPP}\subseteq\class{BQP}$.
Here an advantage of working with the general quantum circuit model
arises: for if one truly believes the Universality Theorem, there is
almost nothing to prove. 

Observe first that the complexity class \class{P} may be defined in
terms of Boolean circuit families in a similar manner to \class{BQP}.
In particular, a given promise problem $A = (A_{\yes},A_{\no})$ is
in \class{P} if and only if there exists a polynomial-time generated
family $C = \{C_n\,:\,n\in\natural\}$ of Boolean circuits, where each
circuit $C_n$ takes $n$ input bits and outputs 1 bit, such that 
\begin{enumerate}
\item[1.]
  $C(x) = 1$ for all $x\in A_{\yes}$, and
\item[2.]
  $C(x) = 0$ for all $x\in A_{\no}$.
\end{enumerate}
A Boolean circuit-based definition of \class{BPP} may be given along similar
lines: a given promise problem $A = (A_{\yes},A_{\no})$ is in \class{BPP}
if and only if there exists a polynomial-bounded function $r$ and a
polynomial-time generated family $C = \{C_n\,:\,n\in\natural\}$ of
Boolean circuits, where each circuit $C_n$ takes $n + r(n)$ input
bits and outputs 1 bit, such that 
\begin{enumerate}
\item[1.]
  $\op{Pr}[C(x,y) = 1] \geq 2/3$ for all $x\in A_{\yes}$, and
\item[2.]
  $\op{Pr}[C(x,y) = 1] \leq 1/3$ for all $x\in A_{\no}$,
\end{enumerate}
where $y\in \Sigma^{r(\abs{x})}$ is chosen uniformly at random in both
cases.

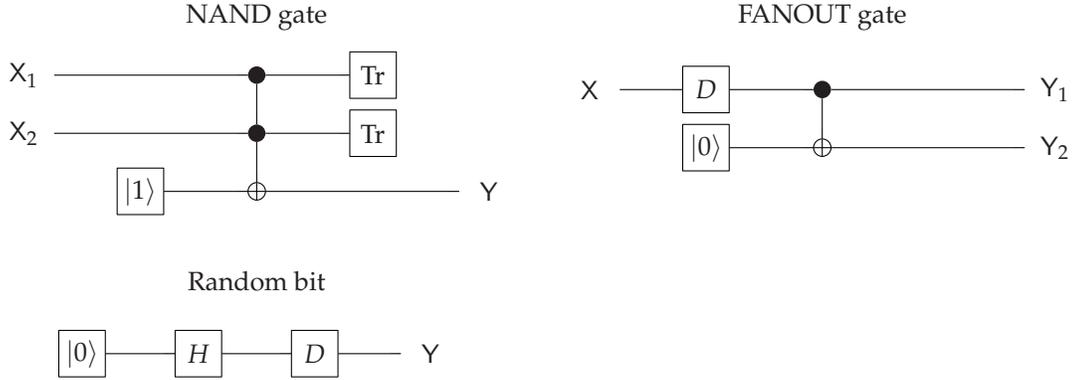
\begin{figure}[t]
  \begin{center}\small
    \unitlength=1.1pt
    \begin{picture}(190, 80)(-20,10)
      \gasset{AHLength=0,ilength=-6}
      
      \node[Nframe=n,Nw=6,Nh=6,fillcolor=Black](T1)(80,60){} 
      \node[Nframe=n,Nw=6,Nh=6,fillcolor=Black](T2)(80,40){} 
      \node[Nw=6,Nh=6](T3)(80,20){} 
      \imark[iangle= 90](T3)
      \imark[iangle= 0](T3) 
      \drawedge(T1,T2){}
      \drawedge(T2,T3){}
      
      \node[Nframe=n](In1)(0,60){$\reg{X}_1$}
      \node[Nframe=n](In2)(0,40){$\reg{X}_2$}
      \node[Nw=16,Nh=16,Nmr=0](1)(40,20){$\ket{1}$}
      \node[Nw=16,Nh=16,Nmr=0](Tr1)(120,60){$\tr$}
      \node[Nw=16,Nh=16,Nmr=0](Tr2)(120,40){$\tr$}
      \node[Nframe=n](Out3)(160,20){$\reg{Y}$}
      
      \drawedge(In1,T1){}
      \drawedge(In2,T2){}
      \drawedge(1,T3){}
      \drawedge(T1,Tr1){}
      \drawedge(T2,Tr2){}
      \drawedge(T3,Out3){}
     
      \put(80,80){\makebox(0,0){NAND gate}}
 
    \end{picture}
    \rule{0cm}{0mm}
    \begin{picture}(190, 80)(-20,5)
      \gasset{AHLength=0,ilength=-6}
      
      \node[Nframe=n,Nw=6,Nh=6,fillcolor=Black](T1)(80,50){} 
      \node[Nw=6,Nh=6](T3)(80,30){} 
      \imark[iangle= 90](T3)
      \imark[iangle= 0](T3) 
      \drawedge(T1,T3){}
      
      \node[Nframe=n](In)(0,50){$\reg{X}$}
      \node[Nw=16,Nh=16,Nmr=0](0)(40,30){$\ket{0}$}
      \node[Nw=16,Nh=16,Nmr=0](D)(40,50){$D$}
      \node[Nframe=n](Out1)(160,50){$\reg{Y}_1$}
      \node[Nframe=n](Out3)(160,30){$\reg{Y}_2$}
      
      \drawedge(In,D){}
      \drawedge(D,T1){}
      \drawedge(0,T3){}
      \drawedge(T1,Out1){}
      \drawedge(T3,Out3){}
     
      \put(80,75){\makebox(0,0){FANOUT gate}}
      
    \end{picture}
    
    \begin{picture}(190, 60)(-40,35)
      \gasset{AHLength=0,ilength=-6}
      
      \node[Nw=16,Nh=16,Nmr=0](0)(0,50){$\ket{0}$}
      \node[Nw=16,Nh=16,Nmr=0](H)(40,50){$H$}
      \node[Nw=16,Nh=16,Nmr=0](D)(80,50){$D$}
      \node[Nframe=n](Out)(120,50){$\reg{Y}$}
      
      \drawedge(0,H){}
      \drawedge(H,D){}
      \drawedge(D,Out){}
     
      \put(60,75){\makebox(0,0){Random bit}}
      
    \end{picture}
    \rule{0cm}{0mm}
    \begin{picture}(190, 60)(-20,30)
    \end{picture}
  \end{center}
  \caption{Quantum circuit implementations of a NAND gate, a FANOUT
    gate, and a random bit.
    The phase-damping gates, denoted by $D$, are only included for
    aesthetic reasons: they force the purely classical behavior that
    would be expected of classical gates, but are not required for the
    quantum simulation of \class{BPP}.}
  \label{fig:nand-fanout-random}
\end{figure}

In both definitions, the circuit family $C$ includes circuits composed of
constant-size Boolean logic gates---which for the sake of brevity 
may be assumed to be composed of NAND gates and FANOUT gates.
(FANOUT operations must be modelled as gates for the sake of the simulation.)
For the randomized case, it may be viewed that the random bits
$y\in\Sigma^{r(\abs{x})}$ are produced by gates that take no input
and output a single uniform random bit.
As NAND gates, FANOUT gates, and random bits are easily implemented
with quantum gates, as illustrated in Figure~\ref{fig:nand-fanout-random},
the circuit family $C$ can be simulated gate-by-gate to
obtain a quantum circuit family $Q = \{Q_n\,:\,n\in\natural\}$
for $A$ that satisfies the definition of \class{BQP}.
It follows that $\class{BPP}\subseteq\class{BQP}$.

\subsection{The \class{BQP} subroutine theorem}

There is an important issue regarding the above definition of
\class{BQP}, which is that it is not an inherently ``clean''
definition with respect to the modularization of algorithms.
The {\it BQP subroutine theorem} of Bennett, Brassard, Bernstein and
Vazirani \cite{BennettBBV97} addresses this issue.

Suppose that it is established that a particular promise problem $A$
is in \class{BQP}, which by definition means that there must exist an
efficient quantum algorithm (represented by a family of quantum
circuits) for $A$.
It is then natural to consider the use of that algorithm as a
subroutine in other quantum algorithms for more complicated problems,
and one would like to be able to do this without worrying
about the specifics of the original algorithm.
Ideally, the algorithm for $A$ should function as an oracle for $A$,
as defined in Section~\ref{sec:oracle-gates}.

A problem arises, however, when queries to an algorithm for $A$ are
made in superposition.
Whereas it is quite common and useful to consider quantum algorithms
that query oracles in superposition, a given \class{BQP} algorithm for
$A$ is only guaranteed to work correctly on classical inputs.
It could be, for instance, that some algorithm for $A$ begins by
applying phase-damping gates to all of its input qubits, or perhaps
this happens inadvertently as a result of the computation.
Perhaps it is too much to ask that the existence of a \class{BQP}
algorithm for $A$ admits a subroutine having the characteristics of an
oracle for $A$?

The \class{BQP} subroutine theorem establishes that, up to
exponentially small error, this is not too much to ask:
the existence of an arbitrary \class{BQP} algorithm for $A$ implies
the existence of a ``clean'' subroutine for $A$ with the
characteristics of an oracle.
A precise statement of the theorem follows.

\begin{theorem}[\class{BQP} subroutine theorem]
  Suppose $A = (A_{\yes},A_{\no})$ is a promise problem in
  \class{BQP}.
  Then for any choice of a polynomial-bounded function $p$ there
  exists a polynomial-bounded function $q$ and a polynomial-time
  generated family of {\it unitary} quantum circuits
  $\{R_n\,:\,n\in\natural\}$ with the following properties:
  \begin{enumerate}
  \item Each circuit $R_n$ implements a unitary operation $U_n$ on 
    $n + q(n) + 1$ qubits.
  \item For every $x\in A_{\yes}$ and $a\in\Sigma$ it holds that
    \[
    \braket{x,0^{m}, a\oplus 1 | U_n | x,0^{m},a} \geq 1 - 2^{-p(n)} 
    \]
    for $n = \abs{x}$ and $m = q(n)$.
  \item For every $x\in A_{\no}$ and $a\in\Sigma$ it holds that
    \[
    \braket{x,0^m,a | U_n | x,0^m,a} \geq 1 - 2^{-p(n)}
    \]
    for $n = \abs{x}$ and $m = q(n)$.
  \end{enumerate}
\end{theorem}

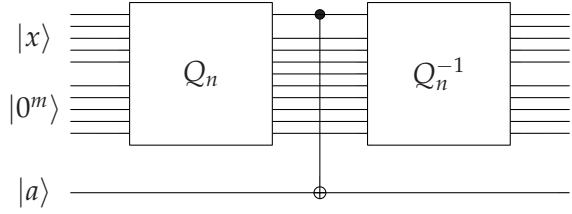
\begin{figure}[t]
  \begin{center}
    \unitlength=0.9pt
    \begin{picture}(300, 100)(0,-10)
      \gasset{AHLength=0,ilength=-5}

      \node[Nw=60,Nh=60,Nmr=0](1)(100,50){$Q_n$}
      \node[Nw=60,Nh=60,Nmr=0](2)(200,50){$Q_n^{-1}$}
      \node[Nw=30,Nh=30,Nmr=0,Nframe=n](In1)(30,65){$\ket{x}$}
      \node[Nw=30,Nh=30,Nmr=0,Nframe=n](In2)(30,35){$\ket{0^m}$}
      \node[Nw=30,Nframe=n](In3)(30,0){$\ket{a}$}

      \node[Nw=30,Nh=30,Nmr=0,Nframe=n](Out1)(270,65){}
      \node[Nw=30,Nh=30,Nmr=0,Nframe=n](Out2)(270,35){}
      \node[Nw=30,Nh=30,Nmr=0,Nframe=n](Out3)(270,0){}

      \node[Nw=4,Nh=4,fillcolor=Black](C1)(150,75){} 

      \node[Nw=5,Nh=5](T1)(150,0){} 
      \imark[iangle= 90](T1)
      \imark[iangle= 0](T1) 
      \drawedge(C1,T1){}
  
      \drawedge[eyo=-25,syo=-25](1,2){}
      \drawedge[eyo=-20,syo=-20](1,2){}
      \drawedge[eyo=-15,syo=-15](1,2){}
      \drawedge[eyo=-10,syo=-10](1,2){}
      \drawedge[eyo=-5,syo=-5](1,2){}
      \drawedge[eyo=0,syo=0](1,2){}
      \drawedge[eyo=5,syo=5](1,2){}
      \drawedge[eyo=10,syo=10](1,2){}
      \drawedge[eyo=15,syo=15](1,2){}

      \drawedge[eyo=25,syo=25](1,2){}

      \drawedge(In3,Out3){}

      \drawedge[eyo=25,syo=10](In1,1){}
      \drawedge[eyo=20,syo=5](In1,1){}
      \drawedge[eyo=15,syo=0](In1,1){}
      \drawedge[eyo=10,syo=-5](In1,1){}
      \drawedge[eyo=5,syo=-10](In1,1){}
      \drawedge[eyo=-5,syo=10](In2,1){}
      \drawedge[eyo=-10,syo=5](In2,1){}
      \drawedge[eyo=-15,syo=0](In2,1){}
      \drawedge[eyo=-20,syo=-5](In2,1){}
      \drawedge[eyo=-25,syo=-10](In2,1){}

      \drawedge[eyo=25,syo=10](Out1,2){}
      \drawedge[eyo=20,syo=5](Out1,2){}
      \drawedge[eyo=15,syo=0](Out1,2){}
      \drawedge[eyo=10,syo=-5](Out1,2){}
      \drawedge[eyo=5,syo=-10](Out1,2){}
      \drawedge[eyo=-5,syo=10](Out2,2){}
      \drawedge[eyo=-10,syo=5](Out2,2){}
      \drawedge[eyo=-15,syo=0](Out2,2){}
      \drawedge[eyo=-20,syo=-5](Out2,2){}
      \drawedge[eyo=-25,syo=-10](Out2,2){}
      
    \end{picture}
  \end{center}

  \caption{A unitary quantum circuit approximating an oracle-gate
    implementation of a \class{BQP} computation. 
    Here $Q_n$ is a unitary purification of a circuit having
    exponentially small error for some problem in \class{BQP}.}
  \label{fig:BQP-subroutine}
\end{figure}

The proof of this theorem is remarkably simple: given a \class{BQP}
algorithm for $A$, one first uses Proposition~\ref{prop:BQP-error} to
obtain a circuit family $Q$ having exponentially small error for $A$.
The circuit illustrated in Figure~\ref{fig:BQP-subroutine} then
implements a unitary operation with the desired properties.
This is essentially a bounded-error quantum adaptation of a classic
construction that allows arbitrary deterministic computations to be
performed reversibly \cite{Bennett73,Toffoli80}.

The following corollary expresses the main implication of the \class{BQP}
subroutine theorem in complexity-theoretic terms.

\begin{cor}
$\class{BQP}^\class{BQP} = \class{BQP}$.
\end{cor}

\subsection{Classical upper bounds on \class{BQP}} \label{sec:GapP}

There is no known way to efficiently simulate quantum computers with
classical computers---and there would be little point in seeking to
build quantum computers if there were.
Nevertheless, some insight into the  limitations of quantum computers
may be gained by establishing containments of \class{BQP} in the
smallest classical complexity classes where this is possible.

The strongest containment known at this time is given by
{\it counting complexity}. 
Counting complexity began with Valiant's work \cite{Valiant79} on the
complexity of computing the permanent, and was further developed and
applied in many papers (including
\cite{BeigelRS95,FennerF+94,Toda91}, among many others).

The basic notion of counting complexity that is relevant to this
article is as follows.
Given a polynomial-time nondeterministic Turing machine $M$ and input
string $x\in\Sigma^{\ast}$, one denotes by $\# M(x)$ the number of
{\it accepting} computation paths of $M$ on $x$, and by
$\#\overline{M}(x)$ the number of {\it rejecting} computation paths of
$M$ on $x$.
A function $f:\Sigma^{\ast} \rightarrow \mathbb{Z}$ is then said to be
a {\it GapP function} if there exists a polynomial-time
nondeterministic Turing machine $M$ such that
$f(x) = \#M(x) - \#\overline{M}(x)$
for all $x\in\Sigma^{\ast}$.

A variety of complexity classes can be specified in terms of
\class{GapP} functions.
For example, a promise problem $A = (A_{\yes},A_{\no})$ is in
\class{PP} if and only if there exists a function $f\in\class{GapP}$
such that $f(x)>0$ for all $x\in A_{\yes}$ and $f(x)\leq 0$ for all
$x\in A_{\no}$.
The remarkable closure properties of \class{GapP} functions allows
many interesting facts to be proved about such classes.
Fortnow's survey \cite{Fortnow97} on counting complexity explains many
of these properties and gives several applications of the theory of
\class{GapP} functions.
The following closure property is used below.

\begin{trivlist}
\item {\bf \class{GapP}--multiplication of matrices.}
Let $p,q:\natural\rightarrow\natural$ be polynomial-bounded functions.
Suppose that for each $n\in\natural$ a sequence of $p(n)$
complex-valued matrices is given:
\[
A_{n,1},\,A_{n,2},\,\ldots,\,A_{n,p(n)},
\]
each having rows and columns indexed by strings in $\Sigma^{q(n)}$.
Suppose further that there exist functions $f,g\in\class{GapP}$ such
that
\begin{align*}
f(1^n,1^k,x,y) & = \op{Re}\left( A_{n,k}[x,y] \right)\\
g(1^n,1^k,x,y) & = \op{Im}\left( A_{n,k}[x,y] \right)
\end{align*}
for all $n\in\natural$, $k\in\{1,\ldots,p(n)\}$, and
$x,y\in\Sigma^{q(n)}$.
Then there exist functions $F,G\in\class{GapP}$ such that
\begin{align*}
F(1^n,x,y) & = \op{Re}\left((A_{n,p(n)}\cdots A_{n,1})[x,y]\right) \\
G(1^n,x,y) & = \op{Im}\left((A_{n,p(n)}\cdots A_{n,1})[x,y]\right)
\end{align*}
for all $n\in\natural$ and $x,y\in\Sigma^{q(n)}$.
In other words, if there exist two \class{GapP} functions
describing the real and imaginary parts of the entries of a polynomial
sequence of matrices, then the same is true of the product of these
matrices.
\end{trivlist}

Now, suppose that $Q = \{Q_n\,:\,n\in\natural\}$ is a polynomial-time
generated family of quantum circuits.
For each $n\in\natural$, let us assume the quantum circuit $Q_n$
consists of gates $G_{n,1},\ldots,G_{n,p(n)}$ for some polynomial
bounded function $p$, labelled in an order that respects the topology
of the circuit.
By tensoring these gates with the identity operation on qubits they do
not touch, and using the natural matrix representation of quantum
operations, it is possible to obtain matrices
$M_{n,1},\ldots,M_{n,p(n)}$ with the property that
\[
M_{n,p(n)} \cdots M_{n,1} \op{vec}(\rho) = \op{vec}(Q_n(\rho))
\]
for every possible input density matrix $\rho$ to the circuit.
The probability that $Q_n$ accepts a given input $x$ is then
\[
\bra{1,1}
\left(M_{n,p(n)} \cdots M_{n,1}\right)
\ket{x,x},
\]
which is a single entry in the product of the matrices.

By padding matrices with rows and columns of zeroes, it may be assumed
that each matrix $M_{n,k}$ has rows and columns indexed by strings of
length $q(n)$ for some polynomial-bounded function $q$.
The assumption that the family $Q$ is polynomial-time generated then
allows one to easily conclude that there exist \class{GapP} functions
$f$ and $g$ so that 
\begin{align*}
f(1^n,1^k,x,y) & = \frac{1}{2}\Re\left( M_{n,k}[x,y] \right)\\
g(1^n,1^k,x,y) & = \frac{1}{2}\Im\left( M_{n,k}[x,y] \right)
\end{align*}
for all $n\in\natural$, $k\in\{1,\ldots,p(n)\}$, and
$x,y\in\Sigma^{q(n)}$.
(Note that his fact makes use of the observation that the natural
matrix representation of all of the gates listed in
Section~\ref{sec:universal-gate-set} have entries whose real and
imaginary parts come from the set $\{-1,-1/2,0,1/2,1\}$.
The numbers $\pm 1/2$ are only needed for the Hadamard gates.)
By the property of \class{GapP} functions above, it follows that there
exists a \class{GapP} function $F$ such that
\[
\op{Pr}[\text{$Q$ accepts $x$}] = \frac{F(x)}{2^{p(\abs{x})}}.
\]

The containment $\class{BQP}\subseteq\class{PP}$ follows easily.
This containment was first proved by Adleman, DeMarrais, and Huang
\cite{AdlemanD+97} using a different method, and was first argued
using counting complexity along similar lines to the above proof by
Fortnow and Rogers \cite{FortnowR99}.
There are two known ways that this upper bound can be improved.
Using the fact that \class{BQP} algorithms have bounded error, Fortnow
and Rogers \cite{FortnowR99} proved that $\class{BQP}$ is contained in
a (somewhat obscure) counting complexity class called \class{AWPP},
which implies the following theorem.

\begin{theorem}
$\class{PP}^{\class{BQP}} = \class{PP}$.
\end{theorem}

\noindent
Another improvement comes from the observation that the above proof
that $\class{BQP}\subseteq\class{PP}$ makes no use of the
bounded-error assumption of \class{BQP}.
It follows that an unbounded error variant of \class{BQP} is equal to
\class{PP}.

\begin{center}
\begin{tabular}{p{0.4in}p{5.7in}}
{\bf \class{PQP}} & 
Let $A = (A_{\yes},A_{\no})$ be a promise problem.
Then $A\in\class{PQP}$ if and only if there exists a polynomial-time
generated family of quantum circuits $Q = \{Q_n\,:\,n\in\natural\}$,
where each circuit $Q_n$ takes $n$ input qubits and produces one
output qubit, that satisfies the following properties.
If $x\in A_{\yes}$ then $\op{Pr}[\text{$Q$ accepts $x$}]>1/2$; and 
if $x\in A_{\no}$ then $\op{Pr}[\text{$Q$ accepts $x$}]\leq 1/2$.
\end{tabular}
\end{center}

\begin{theorem}
$\class{PQP} = \class{PP}$.
\end{theorem}

\subsection{Oracle results involving \class{BQP}}

It is difficult to prove separations among quantum complexity classes 
for apparently the same reasons that this is so for classical
complexity classes.
For instance, one clearly cannot hope to prove 
$\class{BPP} \not= \class{BQP}$ when major collapses such as
$\class{NC} = \class{PP}$ or $\class{P} = \class{PSPACE}$
are still not disproved.
In some cases, however, separations among quantum complexity classes
can be proved in relativized settings, meaning that the separation
holds in the presence of some cleverly defined oracle.
The following oracle results are among several that are known:
\begin{enumerate}
\item 
There exists an oracle $A$ such that
$\class{BQP}^A\not\subseteq\class{MA}^A$
\cite{Babai92, BernsteinV97, Watrous00}.
Such an oracle $A$ intuitively encodes a problem that is solvable
using a quantum computer but is not even efficiently verifiable with a
classical computer.
As the containment $\class{BPP} \subseteq \class{BQP}$ holds relative
to every oracle, this implies that
$\class{BPP}^A \subsetneq \class{BQP}^A$ for this particular choice of $A$.

\item 
There is an oracle $A$ such that
$\class{NP}^A \not\subseteq \class{BQP}^A$ \cite{BennettBBV97}.
In less formal terms: a quantum computer cannot find a needle in an
exponentially large haystack in polynomial time.
This result formalizes a critically important idea, which is that a
quantum computer can only solve a given search problem efficiently if
it is able to exploit that problem's structure.
It is easy to define an oracle search problem represented by $A$ that
has no structure whatsoever, which allows the conclusion 
$\class{NP}^A \not\subseteq \class{BQP}^A$ to be drawn.
It is not currently known whether the \class{NP-complete} problems,
in the absence of an oracle, have enough structure to be solved
efficiently by quantum computers; 
but there is little hope and no indication whatsoever that this should
be so. 
It is therefore a widely believed conjecture that
$\class{NP}\not\subseteq \class{BQP}$.

\item 
  There is an oracle $A$ such that 
  $\class{SZK}^A \not\subseteq \class{BQP}^A$
  \cite{Aaronson02,AaronsonS04}.
  This result is similar in spirit to the previous one, but is
  technically more difficult and rules out the existence of quantum
  algorithms for unstructured collision detection problems.
  The graph isomorphism problem and various problems that arise in
  cryptography are examples of collision detection problems.
  Once again, it follows that quantum algorithms can only solve such
  problems if their structure can be exploited.
  It is a major open question in quantum computing whether the graph
  isomorphism problem is in \class{BQP}.

\end{enumerate}


\section{Quantum proofs}

There are many quantum complexity classes of interest beyond \class{BQP}.
This section concerns one such class, which is a quantum computational
analogue of \class{NP}.
The class is known as \class{QMA}, short for 
{\it quantum Merlin--Arthur}, and is based on the notion of a
{\it quantum proof}:  a quantum state that plays the role of a
certificate or witness to a quantum computer that functions as a
verification procedure.
Interest in both the class \class{QMA} and the general notion of
quantum proofs is primarily based on the fundamental importance of
{\it efficient verification} in computational complexity.
The notion of a quantum proof was first proposed by Knill
\cite{Knill96} and consider more formally by Kitaev 
(presented at a talk in 1999 \cite{Kitaev99} and later published in
\cite{KitaevS+02}).

\subsection{Definition of \class{QMA}}

The definition of \class{QMA} is inspired by the standard definition
of \class{NP} included in Section~\ref{sec:complexity} of this article.
This definition is of course equivalent to the other well-known
definition of \class{NP} based on nondeterministic Turing machines,
but is much better-suited to consideration in the quantum
setting---for nondeterminism is arguably a non-physical notion that
does not naturally extend to quantum computing.
In the definition of \class{NP} from Section~\ref{sec:complexity},
the machine $M$ functions as a {\it verification procedure} that
treats each possible string $y\in\Sigma^{p(\abs{x})}$ as a potential
{\it proof} that $x\in A_{\yes}$.
The conditions on $M$ are known as the {\it completeness} and 
{\it soundness} conditions, which derive their names from logic:
completeness refers to the condition that true statements have proofs,
while soundness refers to the condition that false statements do not.

To define \class{QMA}, the set of possible proofs is extended to
include quantum states, which of course presumes that the verification
procedure is quantum.
As quantum computations are inherently probabilistic, 
a bounded probability of error is allowed in the completeness and
soundness conditions.
(This is why the class is called \class{QMA} rather than \class{QNP},
as it is really \class{MA} and not \class{NP} that is the classical
analogue of \class{QMA}.)

\begin{center}
\begin{tabular}{p{0.4in}p{5.7in}}
{\bf \class{QMA}} & 
Let $A = (A_{\yes},A_{\no})$ be a promise problem, let $p$ be a
polynomial-bounded function, and let $a,b:\natural\rightarrow [0,1]$
be functions.
Then $A\in\class{QMA}_p(a,b)$ if and only if there exists a
polynomial-time generated family of circuits $Q =
\{Q_n\,:\,n\in\natural\}$, where each circuit $Q_n$ takes $n + p(n)$
input qubits and produces one output qubit, with the following properties:
  \begin{enumerate}
  \item {\it Completeness.}
    For all $x\in A_{\yes}$, there exists a $p(\abs{x})$-qubit
    quantum state $\rho$ such that 
    $\op{Pr}[\text{$Q$ accepts $(x, \rho)$}] \geq a(\abs{x})$.
    
  \item {\it Soundness.}
    For all $x\in A_{\no}$ and all $p(\abs{x})$-qubit quantum states
    $\rho$ it holds that 
    $\op{Pr}[\text{$Q$ accepts $(x, \rho)$}] \leq b(\abs{x})$.
  \end{enumerate}
  Also define $\class{QMA} = \bigcup_p \class{QMA}_p(2/3,1/3)$, where
  the union is over all polynomial-bounded functions $p$.
\end{tabular}
\end{center}

\subsection{Problems in \class{QMA}}

Before discussing the general properties of \class{QMA} that are
known, it is appropriate to mention some examples of problems in this
class.
Of course it is clear that thousands of interesting combinatorial
problems are in \class{QMA}, as the containment
$\class{NP}\subseteq\class{QMA}$ is trivial.
What is more interesting is the identification of
problems in \class{QMA} that are not known to be in \class{NP}, for
these are examples that provide insight into the power of quantum
proofs.
There are presently just a handful of known problems in \class{QMA}
that are not known to be in \class{NP}, but the list is growing.

\subsubsection*{The local Hamiltonian problem}

Historically speaking, the first problem identified to be complete for
\class{QMA} was the {\it local Hamiltonian problem}
\cite{Kitaev99,KitaevS+02}, which can be seen as a quantum analogue of
the MAX-$k$-SAT problem. 
Its proof of completeness can roughly be seen as a quantum analogue of
the proof of the Cook--Levin Theorem \cite{Cook72,Levin73}.
The proof has subsequently been strengthened to achieve better
parameters \cite{KempeKR06}.

Suppose that $M$ is a Hermitian matrix whose rows and columns are
indexed by strings of length $n$ for some integer $n\geq 1$.
Then $M$ is said to be {\it $k$-local} if and only if it can be
expressed as
\[
M = P_\pi (A \otimes I) P_\pi^{-1}
\]
for an arbitrary matrix $A$ indexed by $\Sigma^k$, $P_\pi$ a
permutation matrix defined by
\[
P_{\pi} \ket{x_1 \cdots x_n} = \ket{x_{\pi(1)}\cdots x_{\pi(n)}}
\]
for some permutation $\pi\in S_n$, and $I$ denoting the identity
matrix indexed by $\Sigma^{n-k}$.
In less formal terms, $M$ is a matrix that arises from a ``gate'' on
$k$ qubits, but where the gate is described by a $2^k\times 2^k$
Hermitian matrix $A$ rather than a unitary matrix.
It is possible to express such a matrix compactly by specifying $A$ 
along with the bit-positions on which $A$ acts.

Intuitively, a $k$-local matrix assigns a real number (typically
thought of as representing {\it energy}) to any quantum state on $n$
qubits.
This number depends only on the reduced state of the $k$ qubits
where $M$ acts nontrivially, and can be thought of as a locally
defined penalty on a given quantum state.
Loosely speaking, the $k$-local Hamiltonian problem asks whether there
exists a quantum state that can significantly avoid a collection of
such penalties.

\begin{trivlist}
\item
\textsc{The {\em k}-local Hamiltonian problem}
\item
\begin{tabular*}{\textwidth}{@{}l@{\extracolsep{\fill}}p{5.9in}@{}}
{\it Input:} &
A collection $H_1,\ldots,H_m$ of $k$-local Hermitian matrices
with entries indexed by strings of length $n$ and satisfying
$\norm{H_j} \leq 1$ for $j = 1,\ldots,m$. \\[2mm]
{\it Yes:} &
There exists an $n$-qubit quantum state $\ket{\psi}$ such that
$\braket{\psi | H_1 + \cdots + H_m | \psi} \leq -1$.
\\[2mm]
{\it No:} & 
For every $n$-qubit quantum state $\ket{\psi}$ it holds that
$\braket{\psi | H_1 + \cdots + H_m | \psi} \geq 1$.
\end{tabular*}
\end{trivlist}

\begin{theorem}
The 2-local Hamiltonian problem is complete for \class{QMA} with
respect to Karp reductions.
\end{theorem}

The completeness of this problem has been shown to be closely related
to the universality of the so-called {\it adiabatic} model of quantum
computation \cite{AharonovDKLLR07}.

\subsubsection*{Other problems complete for \class{QMA}}

Aside from the local Hamiltonian problem, there are other promise
problems known to be complete for \class{QMA}, including the following:

\begin{enumerate}
\item
Several Variants of the local Hamiltonian problem.
For example, the 2-local Hamiltonian problem remains
\class{QMA}-complete when the local Hamiltonians are restricted to
nearest-neighbor interactions on a two-dimensional array of qubits
\cite{OliveiraT05}.
The hardness of the local Hamiltonian problem with nearest-neighbor
interactions on one-dimensional systems is known to be
\class{QMA}-complete for 12 dimensional particles in place of qubits
\cite{AharonovGIK07}, but is open for smaller systems including qubits.

\item
The {\it density matrix consistency problem}.
Here, the input is a collection of density matrices representing the
reduced states of a hypothetical $n$-qubit state.
The input is a yes-instance of the problem if there exists a state of
the $n$-qubit system that is consistent with the given reduced density
matrices, and is a no-instance if every state of the $n$-qubit system
has reduced states that have significant disagreement from one or more
of the input density matrices.
This problem is known to be complete for \class{QMA} with respect to
Cook reductions \cite{Liu06}, but is not known to be complete for Karp
reductions.
A different problem with similar properties was considered in
\cite{LiuCV07}.

\item
The {\it quantum clique problem}.
Beigi and Shor \cite{BeigiS07} have proved that a promise version of
the following problem is \class{QMA-complete} for any constant $k\geq 2$.
Given a quantum operation $\Phi$, are there $k$ different inputs to
$\Phi$ that are perfectly distinguishable after passing through
$\Phi$?
They name this the quantum clique problem because there is a close
connection between computing the zero-error capacity of a classical
channel and computing the size of the largest clique in the complement
of a graph representing the channel; and this is a quantum variant of
that problem.

\item
Several problems about quantum circuits.
All of the problems mentioned above are proved to be \class{QMA}-hard
through reductions from the local Hamiltonian problem.
Other problems concerning properties of quantum circuits can be
proved \class{QMA-complete} by more direct means, particularly when
the problem itself concerns the existence of a quantum state that
causes an input circuit to exhibit a certain behavior.
An example in this category is the {\it non-identity check} problem 
\cite{JanzingWB05} that asks whether a given unitary circuit
implements an operation that is close to some scalar multiple of the
identity.

\end{enumerate}

\subsubsection*{The group non-membership problem}

Another example of a problem in \class{QMA} that is not known to be
in \class{NP} is the {\it group non-membership problem}
\cite{Babai85,Watrous00}.
This problem is quite different from the \class{QMA}-complete
problems mentioned above in two main respects.
First, the problem corresponds to a language, meaning that the
promise is vacuous: every possible input can be classified as a
yes-instance or no-instance of the problem.
(In all of the above problems it is not possible to do this without
invalidating the known proofs that the problems are in \class{QMA}.)
Second, the problem is not known to be complete for \class{QMA}; and
indeed it would be surprising if \class{QMA} were shown to have
a complete problem having a vacuous promise.

In the group non-membership problem the input is a subgroup $H$ of
some finite group $G$, along with an element $g\in G$.
The yes-instances of the problem are those for which $g\not\in H$,
while the no-instances are those for which $g\in H$.

\begin{trivlist}
\item
\textsc{The group non-membership problem}
\item
\begin{tabular*}{\textwidth}{@{}l@{\extracolsep{\fill}}p{5.9in}@{}}
{\it Input:} &
Group elements $h_1,\ldots,h_k$ and $g$ from some finite group $G$.
Let $H = \langle h_1,\ldots,h_k\rangle$ be the subgroup generated by
$h_1,\ldots,h_k$.
\\[2mm]
{\it Yes:} & $g\not\in H$.\\[2mm]
{\it No:} & $g\in H$.
\end{tabular*}
\end{trivlist}

Like most group-theoretic computational problems, there are many
specific variants of the group non-membership problem that differ in
the way that group elements are represented.
For example, group elements could be represented by permutations in
cycle notation, invertible matrices over a finite field, or any number
of other possibilities.
The difficulty of the problem clearly varies depending the
representation.
The framework of {\it black-box groups}, put forth by Babai and
Szemer{\'e}di \cite{BabaiS84}, simplifies this issue by assuming that
group elements are represented by meaningless labels that can only be
multiplied or inverted by means of a black box, or {\it group oracle}.
Any algorithm or protocol that solves a problem in this framework can
then be adapted to any specific group provided that an efficient
implementation of the group operations exists that can replace the
group oracle.

\begin{theorem}
The group non-membership problem is in \class{QMA} for every choice of
a group oracle.
\end{theorem}

\subsection{Error reduction for \class{QMA}}

The class \class{QMA} is robust with respect to error bounds,
reflected by the completeness and soundness probabilities, in much
the same way that the same is true of $\class{BQP}$.
Two specific error reduction methods are presented in this section:
{\it weak error reduction} and {\it strong error reduction}.
The two methods differ with respect to the length of the quantum proof
that is needed to obtain a given error bound, which is a unique
consideration that arises when working with quantum proofs.

Suppose that $A = (A_{\yes},A_{\no})$ is a given promise problem for
which a \class{QMA}-type quantum verification procedure exists.
To describe both error reduction procedures, it is convenient to
assume that the input to the problem is hard-coded into each circuit
in the family representing the verification procedure.
The verification procedure therefore takes the form
$Q = \{Q_x\,:\,x\in\Sigma^{\ast}\}$, where each circuit $Q_x$ takes a
$p(\abs{x})$-qubit quantum state as input, for some polynomial-bounded
function $p$ representing the quantum proof length, and produces one
output qubit.
The completeness and soundness probabilities are assumed to be given 
by polynomial-time computable functions $a$ and $b$ as usual.

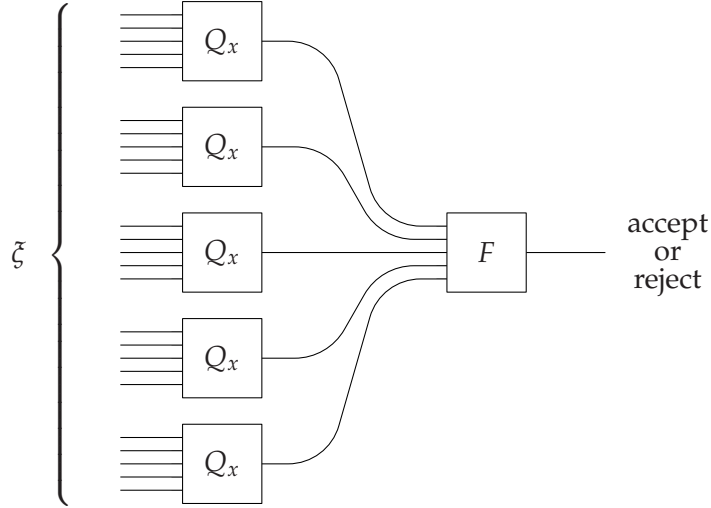
\begin{figure}[t]
  \begin{center}
    \unitlength=1.0pt
    \begin{picture}(200, 180)(20,-10)
      \gasset{AHLength=0}

      \node[Nw=30,Nh=30,Nmr=0](1)(100,160){$Q_x$}
      \node[Nw=30,Nh=30,Nmr=0](2)(100,120){$Q_x$}
      \node[Nw=30,Nh=30,Nmr=0](3)(100,80){$Q_x$}
      \node[Nw=30,Nh=30,Nmr=0](4)(100,40){$Q_x$}
      \node[Nw=30,Nh=30,Nmr=0](5)(100,0){$Q_x$}

      \node[Nframe=n,Nmr=0](In1)(50,160){}
      \node[Nframe=n,Nmr=0](In2)(50,120){}
      \node[Nframe=n,Nmr=0](In3)(50,80){}
      \node[Nframe=n,Nmr=0](In4)(50,40){}
      \node[Nframe=n,Nmr=0](In5)(50,0){}

      \node[Nw=30,Nh=30,Nmr=0](Maj)(200,80){$F$}
      \node[Nframe=n,Nw=50,Nh=30,Nmr=0](Out)(270,80){%
	\makebox(0,0){%
	  \begin{tabular}{c}
	    accept\\[-1mm] or\\[-1mm] reject
	  \end{tabular}
	}}
	
      \drawedge[eyo=10,syo=10](In1,1){}
      \drawedge[eyo=5,syo=5](In1,1){}
      \drawedge[eyo=0,syo=0](In1,1){}
      \drawedge[eyo=-5,syo=-5](In1,1){}
      \drawedge[eyo=-10,syo=-10](In1,1){}
      \drawedge[eyo=10,syo=10](In2,2){}
      \drawedge[eyo=5,syo=5](In2,2){}
      \drawedge[eyo=0,syo=0](In2,2){}
      \drawedge[eyo=-5,syo=-5](In2,2){}
      \drawedge[eyo=-10,syo=-10](In2,2){}
      \drawedge[eyo=10,syo=10](In3,3){}
      \drawedge[eyo=5,syo=5](In3,3){}
      \drawedge[eyo=0,syo=0](In3,3){}
      \drawedge[eyo=-5,syo=-5](In3,3){}
      \drawedge[eyo=-10,syo=-10](In3,3){}
      \drawedge[eyo=10,syo=10](In4,4){}
      \drawedge[eyo=5,syo=5](In4,4){}
      \drawedge[eyo=0,syo=0](In4,4){}
      \drawedge[eyo=-5,syo=-5](In4,4){}
      \drawedge[eyo=-10,syo=-10](In4,4){}
      \drawedge[eyo=10,syo=10](In5,5){}
      \drawedge[eyo=5,syo=5](In5,5){}
      \drawedge[eyo=0,syo=0](In5,5){}
      \drawedge[eyo=-5,syo=-5](In5,5){}
      \drawedge[eyo=-10,syo=-10](In5,5){}
      
      \drawline[arcradius=40](115,160)(140,160)(160,90)(185,90)
      \drawline[arcradius=40](115,120)(140,120)(160,85)(185,85)
      \drawline[arcradius=40](115,80)(140,80)(160,80)(185,80)
      \drawline[arcradius=40](115,40)(140,40)(160,75)(185,75)
      \drawline[arcradius=40](115,0)(140,0)(160,70)(185,70)

      \drawedge(Maj,Out){}
      
      \put(45,80){\makebox(0,0)[r]{$\xi\;\;\left\{\rule{0mm}{35mm}\right.$}}

    \end{picture}
  \end{center}

  \caption{An illustration of the weak error reduction procedure for
    \class{QMA}.
    The circuits $Q_x$ represent the original verification procedure,
    and the circuit labelled $F$ outputs acceptance or rejection based on
    the frequency of accepts among its inputs (which can be adjusted
    depending on $a$ and $b$).
    It cannot be assumed that the input state $\xi$ takes the form of a
    product state $\rho\otimes \cdots\otimes\rho$; but it is not
    difficult to prove that there will always be at least one such 
    state among the states maximizing the acceptance probability of the
    procedure.}
  \label{fig:QMA-error-reduction-original}
\end{figure}

The natural approach to error reduction is repetition: for a given
input $x$ the verification circuit $Q_x$ is evaluated multiple times,
and a decision to accept or reject based on the frequency of accepts
and rejects among the outcomes of the repetitions is made.
Assuming $a(\abs{x})$ and $b(\abs{x})$ are not too close together,
this results in a decrease in error that is exponential in the number
of repetitions \cite{KitaevS+02}.
The problem that arises, however, is that each repetition of the
verification procedure apparently destroys the quantum proof it
verifies, which necessitates the composite verification procedure
receiving $p(\abs{x})$ qubits for {\it each} repetition of the
original procedure as illustrated in
Figure~\ref{fig:QMA-error-reduction-original}.
The form of error reduction implemented by this procedure is called
{\it weak error reduction}, given that the length of the quantum proof
must grow as the error decreases.
(Of course one may view that it also has a strength, which is that it
does not require a significant increase in circuit depth over the
original procedure.)

\begin{figure}[t]
  \begin{center}
    \unitlength=1.0pt
    \begin{picture}(440, 100)(-63,0)
      \gasset{AHLength=0,ilength=-5}

      \node[Nw=30,Nh=60,Nmr=0](1)(20,60){$Q_x$}
      \node[Nw=30,Nh=60,Nmr=0](2)(70,60){$Q_x^{-1}$}
      \node[Nw=30,Nh=60,Nmr=0](3)(120,60){$Q_x$}
      \node[Nw=30,Nh=60,Nmr=0](4)(170,60){$Q_x^{-1}$}
      \node[Nw=30,Nh=60,Nmr=0](5)(220,60){$Q_x$}
      \node[Nw=30,Nh=60,Nmr=0](6)(270,60){$Q_x^{-1}$}

      \node[Nw=30,Nh=30,Nmr=0,Nframe=n](In)(-70,75){$\rho$}
      \node[Nw=30,Nh=30,Nmr=0,Nframe=n](0)(-20,45){}
      \node[Nw=30,Nh=60,Nmr=0](Trash)(320,60){$\tr$}
      \node[Nw=40,Nh=70,Nmr=0](Ancilla)(-25,25){$\ket{0^{k+m}}$}

      \node[Nw=4,Nh=4,fillcolor=Black](C1)(45,85){} 
      \node[Nw=5,Nh=30](C2)(95,45){} 
      \node[Nw=4,Nh=4,fillcolor=Black](C3)(145,85){} 
      \node[Nw=5,Nh=30](C4)(195,45){} 
      \node[Nw=4,Nh=4,fillcolor=Black](C5)(245,85){} 
      \node[Nw=5,Nh=30](C6)(295,45){} 

      \node[Nw=30,Nframe=n](Count1)(-20,20){}
      \node[Nw=30,Nframe=n](Count2)(-20,15){}
      \node[Nw=30,Nframe=n](Count3)(-20,10){}
      \node[Nw=30,Nframe=n](Count4)(-20,5){}
      \node[Nw=30,Nframe=n](Count5)(-20,0){}
      \node[Nw=30,Nframe=n](Count6)(-20,-5){}

      \node[Nw=5,Nh=5](T1)(45,20){} 
      \imark[iangle= 90](T1)
      \imark[iangle= 0](T1) 
      \drawedge(C1,T1){}

      \node[Nw=5,Nh=5](T2)(95,15){} 
      \imark[iangle= 90](T2)
      \imark[iangle= 0](T2) 
      \drawedge(C2,T2){}

      \node[Nw=5,Nh=5](T3)(145,10){} 
      \imark[iangle= 90](T3)
      \imark[iangle= 0](T3) 
      \drawedge(C3,T3){}

      \node[Nw=5,Nh=5](T4)(195,5){} 
      \imark[iangle= 90](T4)
      \imark[iangle= 0](T4) 
      \drawedge(C4,T4){}

      \node[Nw=5,Nh=5](T5)(245,0){} 
      \imark[iangle= 90](T5)
      \imark[iangle= 0](T5) 
      \drawedge(C5,T5){}

      \node[Nw=5,Nh=5](T6)(295,-5){} 
      \imark[iangle= 90](T6)
      \imark[iangle= 0](T6) 
      \drawedge(C6,T6){}
      
      \node[Nw=30,Nh=30,Nmr=0](Total)(320,7.5){$C$}
      \node[Nw=40,Nframe=n](Output)(370,7.5){output}
      
      \drawedge(Count1,T1){}
      \drawedge(Count2,T2){}
      \drawedge(Count3,T3){}
      \drawedge(Count4,T4){}
      \drawedge(Count5,T5){}
      \drawedge(Count6,T6){}

      \drawedge(Total,Output){}

      \drawedge[eyo=12.5](T1,Total){}
      \drawedge[eyo=7.5](T2,Total){}
      \drawedge[eyo=2.5](T3,Total){}
      \drawedge[eyo=-2.5](T4,Total){}
      \drawedge[eyo=-7.5](T5,Total){}
      \drawedge[eyo=-12.5](T6,Total){}

      \drawedge[syo=25](1,C1){}
      \drawedge[eyo=25](C1,2){}
      \drawedge[syo=25](3,C3){}
      \drawedge[eyo=25](C3,4){}
      \drawedge[syo=25](5,C5){}
      \drawedge[eyo=25](C5,6){}


      \drawedge[eyo=-25,syo=-25](1,2){}
      \drawedge[eyo=-20,syo=-20](1,2){}
      \drawedge[eyo=-15,syo=-15](1,2){}
      \drawedge[eyo=-10,syo=-10](1,2){}
      \drawedge[eyo=-5,syo=-5](1,2){}
      \drawedge[eyo=0,syo=0](1,2){}
      \drawedge[eyo=5,syo=5](1,2){}
      \drawedge[eyo=10,syo=10](1,2){}
      \drawedge[eyo=15,syo=15](1,2){}


      \drawedge[eyo=-25,syo=-25](3,4){}
      \drawedge[eyo=-20,syo=-20](3,4){}
      \drawedge[eyo=-15,syo=-15](3,4){}
      \drawedge[eyo=-10,syo=-10](3,4){}
      \drawedge[eyo=-5,syo=-5](3,4){}
      \drawedge[eyo=0,syo=0](3,4){}
      \drawedge[eyo=5,syo=5](3,4){}
      \drawedge[eyo=10,syo=10](3,4){}
      \drawedge[eyo=15,syo=15](3,4){}


      \drawedge[eyo=-25,syo=-25](5,6){}
      \drawedge[eyo=-20,syo=-20](5,6){}
      \drawedge[eyo=-15,syo=-15](5,6){}
      \drawedge[eyo=-10,syo=-10](5,6){}
      \drawedge[eyo=-5,syo=-5](5,6){}
      \drawedge[eyo=0,syo=0](5,6){}
      \drawedge[eyo=5,syo=5](5,6){}
      \drawedge[eyo=10,syo=10](5,6){}
      \drawedge[eyo=15,syo=15](5,6){}


      \drawedge[eyo=-10,syo=-25](2,C2){}
      \drawedge[eyo=-5,syo=-20](2,C2){}
      \drawedge[eyo=0,syo=-15](2,C2){}
      \drawedge[eyo=5,syo=-10](2,C2){}
      \drawedge[eyo=10,syo=-5](2,C2){}

      \drawedge[eyo=-10,syo=-25](3,C2){}
      \drawedge[eyo=-5,syo=-20](3,C2){}
      \drawedge[eyo=0,syo=-15](3,C2){}
      \drawedge[eyo=5,syo=-10](3,C2){}
      \drawedge[eyo=10,syo=-5](3,C2){}
      
      \drawedge[eyo=5,syo=5](2,3){}
      \drawedge[eyo=10,syo=10](2,3){}
      \drawedge[eyo=15,syo=15](2,3){}
      \drawedge[eyo=20,syo=20](2,3){}
      \drawedge[eyo=25,syo=25](2,3){}


      \drawedge[eyo=-10,syo=-25](4,C4){}
      \drawedge[eyo=-5,syo=-20](4,C4){}
      \drawedge[eyo=0,syo=-15](4,C4){}
      \drawedge[eyo=5,syo=-10](4,C4){}
      \drawedge[eyo=10,syo=-5](4,C4){}

      \drawedge[eyo=-10,syo=-25](5,C4){}
      \drawedge[eyo=-5,syo=-20](5,C4){}
      \drawedge[eyo=0,syo=-15](5,C4){}
      \drawedge[eyo=5,syo=-10](5,C4){}
      \drawedge[eyo=10,syo=-5](5,C4){}
      
      \drawedge[eyo=5,syo=5](4,5){}
      \drawedge[eyo=10,syo=10](4,5){}
      \drawedge[eyo=15,syo=15](4,5){}
      \drawedge[eyo=20,syo=20](4,5){}
      \drawedge[eyo=25,syo=25](4,5){}


      \drawedge[eyo=-10,syo=-25](6,C6){}
      \drawedge[eyo=-5,syo=-20](6,C6){}
      \drawedge[eyo=0,syo=-15](6,C6){}
      \drawedge[eyo=5,syo=-10](6,C6){}
      \drawedge[eyo=10,syo=-5](6,C6){}

      \drawedge[eyo=-10,syo=-25](Trash,C6){}
      \drawedge[eyo=-5,syo=-20](Trash,C6){}
      \drawedge[eyo=0,syo=-15](Trash,C6){}
      \drawedge[eyo=5,syo=-10](Trash,C6){}
      \drawedge[eyo=10,syo=-5](Trash,C6){}
      
      \drawedge[eyo=5,syo=5](6,Trash){}
      \drawedge[eyo=10,syo=10](6,Trash){}
      \drawedge[eyo=15,syo=15](6,Trash){}
      \drawedge[eyo=20,syo=20](6,Trash){}
      \drawedge[eyo=25,syo=25](6,Trash){}


      \drawedge[eyo=25,syo=10](In,1){}
      \drawedge[eyo=20,syo=5](In,1){}
      \drawedge[eyo=15,syo=0](In,1){}
      \drawedge[eyo=10,syo=-5](In,1){}
      \drawedge[eyo=5,syo=-10](In,1){}
      
      \drawedge[eyo=-5,syo=10](0,1){}
      \drawedge[eyo=-10,syo=5](0,1){}
      \drawedge[eyo=-15,syo=0](0,1){}
      \drawedge[eyo=-20,syo=-5](0,1){}
      \drawedge[eyo=-25,syo=-10](0,1){}

    \end{picture}
  \end{center}

  \caption{Illustration of the strong error reduction procedure for
    \class{QMA}.
    The circuit $Q_x$ represents a unitary purification of a given
    \class{QMA} verification procedure on an input $x$, while the
    circuit $C$ determines whether to accept or reject based on the
    number of alternations of its input qubits.
    The quantum proof is denoted by $\rho$.}
  \label{fig:strong-error-reduction}
\end{figure}
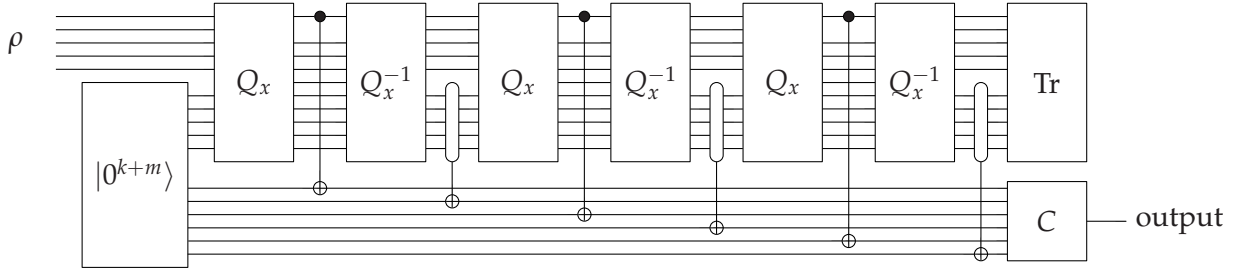

The second error reduction procedure for \class{QMA} was described in
\cite{MarriottW05}.
Like weak error reduction it gives an exponential reduction in error
for roughly a linear increase in circuit size, but has the advantage
that it does not require any increase in the length of the quantum
proof.
This form of error reduction is called {\it strong error reduction}
because of this advantage, which turns out to be quite handy in some
situations.
The procedure is illustrated in Figure~\ref{fig:strong-error-reduction}.
The following theorem follows from an analysis of this procedure.

\begin{theorem}
  \label{theorem:QMA-error-reduction}
  Suppose that $a,b:\natural \rightarrow [0,1]$ are polynomial-time
  computable functions and $q:\natural\rightarrow\natural$ is a
  polynomial-bounded function such that
  $a(n) - b(n) \geq 1/q(n)$ for all but finitely many $n\in\natural$.
  Then for every choice of polynomial-bounded functions
  $p,r:\natural\rightarrow\natural$ such that $r(n) \geq 2$ for all
  but finitely many $n$, it holds that
  \[
  \class{QMA}_p(a,b) = \class{QMA}_p\left(1 - 2^{-r}, 2^{-r}\right).
  \]
\end{theorem}

\subsection{Containment of \class{QMA} in \class{PP}}

By means of strong error reduction for \class{QMA}, it can be
proved that \class{QMA} is contained in \class{PP} as follows.
Suppose that some promise problem $A = (A_{\yes},A_{\no})$ is
contained in \class{QMA}.
Then by Theorem~\ref{theorem:QMA-error-reduction} it holds that
\begin{equation} \label{eq:small-error-QMA}
A\in\class{QMA}_p\left(1 - 2^{-(p+2)}, 2^{-(p+2)}\right)
\end{equation}
for some polynomial-bounded function $p$.
What is important here is that the soundness probability is smaller
than the reciprocal of the dimension of the space corresponding to the
quantum proof (which is why strong error reduction is needed).

Now, consider an algorithm that does not receive any quantum proof,
but instead just randomly guesses a quantum proof on $p$ qubits
and feeds this proof into a verification procedure having
completeness and soundness probabilities consistent with the above
inclusion \eqref{eq:small-error-QMA}.
To be more precise, the quantum proof is substituted by the totally
mixed state on $p$ qubits.
A simple analysis shows that this algorithm accepts every string
$x\in A_{\yes}$ with probability at least $2^{-(p(\abs{x})+1)}$ and
accepts every string $x\in A_{\no}$ with probability at most
$2^{-(p(\abs{x})+2)}$.
Both of these probabilities are exponentially small, but the
separation between them is enough to establish that
$A\in\class{PQP} = \class{PP}$.

\subsection{Classical proofs for quantum verification procedures}

One may also consider quantum verification procedures that receive
classical proofs rather than quantum proofs.
Aharonov and Naveh \cite{AharonovN02} defined a complexity class
\class{MQA} accordingly.

\begin{center}
\begin{tabular}{p{0.5in}p{5.6in}}
{\bf \class{MQA}} & 
Let $A = (A_{\yes},A_{\no})$ be a promise problem.
Then $A\in\class{MQA}$ if and only if there exists a
polynomial-bounded function $p$ and a polynomial-time generated family
of circuits $Q = \{Q_n\,:\,n\in\natural\}$, where each circuit $Q_n$
takes $n + p(n)$ input qubits and produces one output qubit, with the
following properties.
For all $x\in A_{\yes}$, there exists a string
$y\in\Sigma^{p(\abs{x})}$ such that
$\op{Pr}[\text{$Q$ accepts $(x, y)$}] \geq 2/3$;
and for all $x\in A_{\no}$ and all strings $y\in\Sigma^{p(\abs{x})}$
it holds that $\op{Pr}[\text{$Q$ accepts $(x, y)$}] \leq 1/3$.
\end{tabular}
\end{center}

\noindent
(This class was originally named \class{QCMA}, and is more commonly
known by that name---but it is not too late to give this interesting
class a better name.)

When considering the power of quantum proofs, it is the question of
whether \class{MQA} is properly contained in \class{QMA} that arguably
cuts to the heart of the issue.
Aaronson and Kuperberg \cite{AaronsonK07} studied this question, and
based on a reasonable group-theoretic conjecture argued that
the group non-membership problem is likely to be in \class{MQA}.

\subsection{Are two quantum proofs better than one?}

An unusual, yet intriguing question about quantum proofs was asked by
Kobayashi, Matsumoto, and Yamakami \cite{KobayashiM+03}:
{\it are two quantum proofs better than one?}
The implicit setting in this question is similar to that of
\class{QMA}, except that the verification procedure receives {\it two}
quantum proofs that are {\it guaranteed to be unentangled}.
The complexity class $\class{QMA(2)}$ is defined to be the class of
promise problems having such two-proof systems with completeness and
soundness probabilities $2/3$ and $1/3$, respectively.
(It is not known to what extent this type of proof system is robust
with respect to error bounds, so it is conceivable that other
reasonable choices of error bounds could give distinct complexity
classes.)
The class $\class{QMA(2)}$, and its natural extension to more than two
proofs, have subsequently been studied in \cite{AaronsonBDFS08,BlierT07}.

The restriction on entanglement that is present in the definition of
\class{QMA(2)} may seem artificial from a physical perspective, as there
is no obvious mechanism that would prevent two entities with otherwise
unlimited computational power from sharing entanglement.
Nevertheless, the question of the power of $\class{QMA(2)}$ is
interesting in that it turns around the familiar question in quantum
computing about the computational power of entanglement.
Rather than asking if computational power is limited by a constraint
on entanglement, here the question is whether a constraint on
entanglement {\it enhances} computational power.
It is clear that two quantum proofs are no less powerful than one, as
a verification procedure may simply ignore one of the proofs---which
means that $\class{QMA}\subseteq\class{QMA(2)}$.
Good upper bounds on $\class{QMA(2)}$, however, are not known: the best
upper bound presently known is $\class{QMA(2)} \subseteq
\class{NEXP}$, which follows easily from the fact that a nondeterministic
exponential-time algorithm can guess explicit descriptions of the
quantum proofs and then perform an exponential-time simulation of the
verification procedure.

\section{Quantum interactive proof systems}

The notion of efficient proof verification is generalized by means of
the {\it interactive proof system} model, which has fundamental
importance in complexity theory and theoretical cryptography.
Interactive proof systems, or {\it interactive proofs} for short, were
first introduced by Babai \cite{Babai85,BabaiM88} and Goldwasser,
Micali, and Rackoff \cite{GoldwasserMR85,GoldwasserMR89}, and have led to
remarkable discoveries in complexity such as the {\it PCP Theorem}
\cite{AroraLMSS98,AroraS98,Dinur07}.
In the most commonly studied variant of interactive proof systems, a
polynomial-time {\it verifier} interacts with a computationally
unbounded {\it prover} that tries to convince the verifier of the
truth of some statement.
The prover is not trustworthy, however, and so the verifier must be
specified in such a way that it is not convinced that false statements
are true.

{\it Quantum interactive proof systems}
\cite{Watrous03-pspace,KitaevW00}
are interactive proof systems in which the prover and verifier may
exchange and process quantum information.
This ability of both the prover and verifier to exchange and process
quantum information endows quantum interactive proofs with interesting
properties that distinguish them from classical interactive proofs, and
illustrate unique features of quantum information.

\subsection{Definition of quantum interactive proofs and the class
  \class{QIP}}

\begin{figure}[t]
  \begin{center}
    \unitlength=1.9pt
    \begin{picture}(225, 60)(0,0)

      \node[Nframe=n,Nw=10](In1)(10,40){$x$}
      \node[Nframe=n,Nw=10](In2)(10,5){$x$}

      \node[Nframe=n,Nw=24](Out)(220,40){%
	\makebox(0,0){%
	  \begin{tabular}{c}
	    accept\\[-1mm] or\\[-1mm] reject
	  \end{tabular}
	}}
      
      \node[Nw=10,Nh=20,Nmr=0](V1)(35,40){$V_1$}
      \node[Nw=10,Nh=20,Nmr=0](V2)(85,40){$V_2$}
      \node[Nw=10,Nh=20,Nmr=0](V3)(135,40){$V_3$}
      \node[Nw=10,Nh=20,Nmr=0](V4)(185,40){$V_4$}

      \node[Nw=10,Nh=20,Nmr=0](P1)(60,10){$P_1$}
      \node[Nw=10,Nh=20,Nmr=0](P2)(110,10){$P_2$}
      \node[Nw=10,Nh=20,Nmr=0](P3)(160,10){$P_3$}

      \drawbpedge[syo=-5,eyo=5](V1,24,24,P1,24,-24){1} 
      \drawbpedge[syo=-5,eyo=5](V2,24,24,P2,24,-24){3} 
      \drawbpedge[syo=-5,eyo=5](V3,24,24,P3,24,-24){5} 

      \drawbpedge[syo=5,eyo=-5](P1,-24,24,V2,-24,-24){2} 
      \drawbpedge[syo=5,eyo=-5](P2,-24,24,V3,-24,-24){4} 
      \drawbpedge[syo=5,eyo=-5](P3,-24,24,V4,-24,-24){6} 
      
      \drawedge[syo=5,eyo=5](V1,V2){}
      \drawedge[syo=5,eyo=5](V2,V3){}
      \drawedge[syo=5,eyo=5](V3,V4){}

      \drawedge[syo=-5,eyo=-5](P1,P2){}
      \drawedge[syo=-5,eyo=-5](P2,P3){}
      
      \drawedge(In1,V1){}
      \drawedge[eyo=-5](In2,P1){}
      \drawedge(V4,Out){}
    \end{picture}
  \end{center}

  \caption{A quantum interactive proof system.  There are six
    messages in this example, labelled $1,\ldots,6$.
    (There may be polynomially many messages in general.)
    The arrows each represent a collection of qubits, rather than single
    qubits as in previous figures.
    The superscript $n$ is omitted in the names of the prover and
    verifier circuits (which can safely be done when the input length
    $n$ is determined by context).}
  \label{fig:quantum-interactive-proof}
\end{figure}
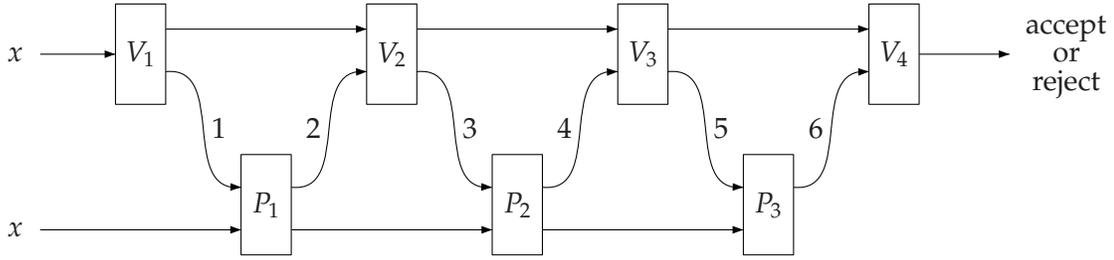

A quantum interactive proof system involves an interaction between a
prover and verifier as suggested by 
Figure~\ref{fig:quantum-interactive-proof}.

The verifier is described by a polynomial-time generated family
\[
V = \left\{
V^n_j\,:\,n\in\natural,\:\,j \in \{1,\ldots,p(n)\}\right\}
\]
of quantum circuits, for some polynomial-bounded function $p$.
On an input string $x$ having length~$n$, the sequence of circuits
$V^n_1,\ldots,V^n_{p(n)}$
determine the actions of the verifier over the course of the
interaction, with the value $p(n)$ representing the number of 
{\it turns} the verifier takes.
For instance, $p(n) = 4$ in the example depicted in
Figure~\ref{fig:quantum-interactive-proof}.
The inputs and outputs of the verifier's circuits are divided into two
categories: private {\it memory} qubits and {\it message} qubits.
The message qubits are sent to, or received from, the prover, while
the memory qubits are retained by the verifier as illustrated in
the figure.
It is always assumed that the verifier {\it receives} the last
message; so if the number of messages is to be $m = m(n)$, then it
must be the case that $p(n) = \left\lfloor m/2\right\rfloor + 1$.

The prover is defined in a similar way to the verifier, although no
computational assumptions are made: the prover is a family of arbitrary
quantum operations 
\[
P = \left\{
P^n_j\,:\,n\in\natural,\:\,j \in \{1,\ldots,q(n)\}\right\}
\]
that interface with a given verifier in the natural way (assuming the
number and sizes of the message spaces are compatible).
Again, this is as suggested by
Figure~\ref{fig:quantum-interactive-proof}.

Now, on a given input string $x$, the prover $P$ and verifier $V$ have
an interaction by composing their circuits as described above, after
which the verifier measures an output qubit to determine acceptance or
rejection.
In direct analogy to the classes \class{NP}, \class{MA}, and
\class{QMA}, one defines quantum complexity classes based on the
completeness and soundness properties of such interactions.

\begin{center}
\begin{tabular}{p{0.4in}p{5.5in}}
{\bf \class{QIP}} & 
Let $A = (A_{\yes},A_{\no})$ be a promise problem, let $m$ be a
polynomial-bounded function, and let $a,b:\natural\rightarrow [0,1]$
be polynomial-time computable functions.
Then $A\in\class{QIP}(m,a,b)$ if and only if there exists an
$m$-message quantum verifier $V$ with the following properties:
\begin{enumerate}
\item {\it Completeness.}
  For all $x\in A_{\yes}$, there exists a quantum prover $P$ that
  causes $V$ to accept $x$ with probability at least $a(\abs{x})$.
  
\item {\it Soundness.}
  For all $x\in A_{\no}$, every quantum prover $P$ causes $V$ to
  accept $x$ with probability at most $b(\abs{x})$.
  
\end{enumerate}
Also define $\class{QIP}(m) = \class{QIP}(m,2/3,1/3)$ for each
polynomial-bounded function $m$ and define
$\class{QIP} = \bigcup_m \class{QIP}(m)$, where the union is over all
polynomial-bounded functions $m$.
\end{tabular}
\end{center}

\subsection{Properties of quantum interactive proofs}

Classical interactive proofs can trivially be simulated by
quantum interactive proofs, and so the containment
$\class{PSPACE}\subseteq\class{QIP}$ follows directly from
$\class{IP}=\class{PSPACE}$ \cite{LundF+92,Shamir92}.
Unlike classical interactive proofs, quantum interactive proofs are
not known to be simulatable in \class{PSPACE}.
Simulation is possible in \class{EXP} through the use of semidefinite
programming \cite{KitaevW00}.

\begin{theorem}
$\class{QIP}\subseteq\class{EXP}$.
\end{theorem}

Quantum interactive proofs, like ordinary classical interactive
proofs, are quite robust with respect to the choice of completeness
and soundness probabilities. 
In particular, the following facts hold \cite{KitaevW00,GutoskiW07}.

\begin{enumerate}
\item
Every quantum interactive proof can be transformed into
an equivalent quantum interactive proof with {\it perfect completeness}:
the completeness probability is 1 and the soundness probability is
bounded away from 1.
The precise bound obtained for the soundness probability depends on
the completeness and soundness probabilities of the original proof
system. 
This transformation to a perfect-completeness proof system comes at
the cost of one additional round (i.e., two messages) of communication.

\item
Parallel repetition of quantum interactive proofs with perfect
completeness gives an exponential reduction in soundness error.
An exponential reduction in completeness and soundness error is
also possible for quantum interactive proofs not having perfect
completeness, but the procedure is slightly more complicated than for
the perfect completeness case.
\end{enumerate}

One of the major differences between quantum and classical interactive
proof systems is that quantum interactive proofs can be 
{\it parallelized}:
every quantum interactive proof system, possibly having polynomially
many rounds of communication, can be transformed into an
equivalent quantum interactive proof system with {\it three messages}
\cite{KitaevW00}.
This transformation comes at the cost of weakening the error bounds.
However, it preserves perfect completeness, and the soundness error
can subsequently be reduced by parallel repetition without increasing
the number of messages beyond three.
If classical interactive proof systems could be parallelized in this
way, then it would follow that $\class{AM}=\class{PSPACE}$; a collapse
that would surprise most complexity theorists and have major
implications to the theory.

The following theorem summarizes the implications of the facts just
discussed to the complexity classes $\class{QIP}(m,a,b)$ defined
above.

\begin{theorem}\label{theorem:QIP-main}
  Let $a,b:\natural\rightarrow [0,1]$ be polynomial-time computable
  functions and let $p$ be a polynomial-bounded function such that
  $a(n) - b(n) \geq \frac{1}{p(n)}$
  for all but finitely many $n\in\natural$.
  Also let $m$ and $r$ be polynomial-bounded functions.
  Then $\class{QIP}(m,a,b) \subseteq \class{QIP}(3,1,2^{-r})$.
\end{theorem}

For a wide range of completeness and soundness probabilities
this leaves just four complexity classes among those defined above:
$\class{QIP}(0) = \class{BQP}$, $\class{QIP}(1) = \class{QMA}$,
$\class{QIP}(2)$, and $\class{QIP}(3) = \class{QIP}$.

The final property of quantum interactive proofs that will be
discussed in this section is the existence of an interesting complete
promise problem \cite{RosgenW05}.
The problem is to determine whether the operations induced by two
quantum circuits are significantly different, or are approximately the
same, with respect to the same metric on quantum operations discussed
in Section~\ref{sec:universal-gate-set}.

\begin{trivlist}
\item
\textsc{The quantum circuit distinguishability problem}
\item
\begin{tabular*}{\textwidth}{@{}l@{\extracolsep{\fill}}p{5.9in}@{}}
{\it Input:} & 
Quantum circuits $Q_0$ and $Q_1$, both taking $n$
input qubits and producing $m$ output qubits.\\[2mm]
{\it Yes:} &
$\delta(Q_0,Q_1) \geq 2/3$.\\[2mm]
{\it No:} & 
$\delta(Q_0,Q_1) \leq 1/3$.
\end{tabular*}
\end{trivlist}

\begin{theorem}
The quantum circuit distinguishability problem is \class{QIP}-complete
with respect to Karp reductions.
\end{theorem}

\subsection{Zero-knowledge quantum interactive proofs}

Interactive proof systems can sometimes be made to have a
cryptographically motivated property known as the {\it zero-knowledge}
property \cite{GoldwasserMR89}.
Informally speaking, an interactive proof system is zero-knowledge if
a verifier ``learns nothing'' from an interaction with the prover on
an input $x\in A_{\yes}$, beyond the fact that it is indeed the case
that $x\in A_{\yes}$.
This should hold even if the verifier deviates from the actions
prescribed to it by the interactive proof being considered.
At first this notion seems paradoxical, but nevertheless there are
many interesting examples of such proof systems.
For an introductory survey on zero-knowledge, see \cite{Vadhan07}.

Several variants of zero-knowledge are often studied in the classical
setting that differ in the particular way that the notion of
``learning nothing'' is formalized.
This article will only consider {\it statistical} zero-knowledge,
which is the variant of zero-knowledge that is most easily adapted to the
quantum setting.

\begin{figure}[t]
  \begin{center}
    \unitlength=1.9pt
    \begin{picture}(225, 60)(-3,0)

      \node[Nframe=n,Nw=15](VIn1)(-10,45){$\rho$}
      \node[Nframe=n,Nw=10](VIn2)(15,35){$x$}
      \node[Nframe=n,Nw=10](PIn2)(15,5){$x$}
      \node[Nframe=n,Nw=20](Out)(225,45){$\Phi_x(\rho)$}
      
      \node[Nw=10,Nh=20,Nmr=0](V1)(35,40){$V'_1$}
      \node[Nw=10,Nh=20,Nmr=0](V2)(85,40){$V'_2$}
      \node[Nw=10,Nh=20,Nmr=0](V3)(135,40){$V'_3$}
      \node[Nw=10,Nh=20,Nmr=0](V4)(185,40){$V'_4$}

      \node[Nw=10,Nh=20,Nmr=0](P1)(60,10){$P_1$}
      \node[Nw=10,Nh=20,Nmr=0](P2)(110,10){$P_2$}
      \node[Nw=10,Nh=20,Nmr=0](P3)(160,10){$P_3$}
      
      \drawbpedge[syo=-5,eyo=5](V1,24,24,P1,24,-24){} 
      \drawbpedge[syo=-5,eyo=5](V2,24,24,P2,24,-24){} 
      \drawbpedge[syo=-5,eyo=5](V3,24,24,P3,24,-24){} 

      \drawbpedge[syo=5,eyo=-5](P1,-24,24,V2,-24,-24){} 
      \drawbpedge[syo=5,eyo=-5](P2,-24,24,V3,-24,-24){} 
      \drawbpedge[syo=5,eyo=-5](P3,-24,24,V4,-24,-24){} 
      
      \drawedge[syo=5,eyo=5](V1,V2){}
      \drawedge[syo=5,eyo=5](V2,V3){}
      \drawedge[syo=5,eyo=5](V3,V4){}

      \drawedge[syo=-5,eyo=-5](P1,P2){}
      \drawedge[syo=-5,eyo=-5](P2,P3){}
      
      \drawedge[eyo=5](VIn1,V1){}
      \drawedge[eyo=-5](VIn2,V1){}
      \drawedge[eyo=-5](PIn2,P1){}
      \drawedge[syo=5](V4,Out){}
    \end{picture}
  \end{center}

  \caption{A cheating verifier $V'$ performs a quantum operation
    $\Phi_x$ with the unintentional help of the honest prover.}
  \label{fig:quantum-zero-knowledge}
\end{figure}
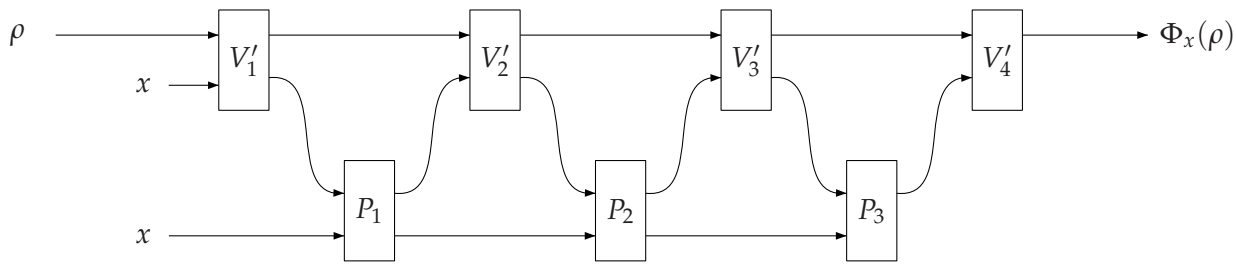

Suppose that $A = (A_{\yes},A_{\no})$ is a promise problem, and
$(V,P)$ is a quantum interactive proof system for $A$.
By this it is meant that $V$ is the {\it honest verifier} and $P$ is
the {\it honest prover} that behave precisely as the proof system
specifies.
Whether or not this proof system possesses the zero-knowledge property
depends on the characteristics of interactions between a 
{\it cheating verifier} $V'$ with the honest prover $P$ on inputs 
$x\in A_{\yes}$.
Figure~\ref{fig:quantum-zero-knowledge} illustrates such an
interaction, wherein it is viewed that the cheating verifier $V'$ is
using the interaction with $P$ on input $x$ to compute some quantum
operation $\Phi_x$.
Informally speaking, the input to this operation represents the
verifier's state of knowledge before the protocol is run, while the
output represents the verifier's state of knowledge after.

Now, the proof system $(V,P)$ is said to be 
{\it quantum statistical zero-knowledge} if, for any choice of a
polynomial-time cheating verifier $V'$, the quantum operation $\Phi_x$
can be efficiently approximated for all $x\in A_{\yes}$.
More precisely, the assumption that $V'$ is described by
polynomial-time generated quantum circuits must imply that there
exists a polynomial-time generated family
$\{Q_x\,:\,x\in\Sigma^{\ast}\}$ of quantum circuits for which
$\delta(\Phi_x,Q_x)$ is negligible for every $x\in A_{\yes}$. 
Intuitively this definition captures the notion of learning
nothing---for anything that the cheating verifier could compute in
polynomial time with the help of the prover could equally well have
been computed in polynomial time without the prover's help.

The class of promise problems having statistical zero-knowledge
quantum interactive proofs is denoted \class{QSZK}.

\begin{center}
\begin{tabular}{p{0.6in}p{5.5in}}
{\bf \class{QSZK}} & A promise problem $A = (A_{\yes},A_{\no})$ is in
\class{QSZK} if and only if it has a statistical zero-knowledge quantum
interactive proof system.
\end{tabular}
\end{center}

\noindent
Although it has not been proved, it is reasonable to conjecture that
\class{QSZK} is properly contained in \class{QIP}; for the
zero-knowledge property seems to be quite restrictive.
Indeed, it was only recently established that there exist non-trivial
quantum interactive proof systems that are statistical zero-knowledge
\cite{Watrous06}.
The following facts \cite{Watrous02,Watrous06} are among those known
about \class{QSZK}.

\begin{enumerate}

\item
Statistical zero-knowledge quantum interactive proof systems can be
parallelized to two messages.
It follows that $\class{QSZK} \subseteq \class{QIP}(2)$.

\item
The class \class{QSZK} is closed under complementation:
a given promise problem $A$ has a quantum statistical zero-knowledge proof
if and only if the same is true for the problem obtained by exchanging
the yes- and no-instances of $A$.

\item
Statistical zero-knowledge quantum interactive proof systems can be
simulated in polynomial space:
$\class{QSZK} \subseteq \class{PSPACE}$.

\end{enumerate}

\noindent
Classical analogues to the first and second facts in this list were
shown first \cite{SahaiV03}.
(The classical analogue to the third fact is
$\class{SZK}\subseteq\class{PSPACE}$, which follows trivially from 
$\class{IP}\subseteq\class{PSPACE}$.)
A key step toward proving the above properties (which is also similar
to the classical case) is to establish that the following promise problem
is complete for \class{QSZK}.
The problem is a restricted version of the \class{QIP}-complete
quantum circuit distinguishability problem.

\pagebreak[3]

\begin{trivlist}
\item
\textsc{The quantum state distinguishability problem}
\item
\begin{tabular*}{\textwidth}{@{}l@{\extracolsep{\fill}}p{5.9in}@{}}
{\it Input:} & 
Quantum circuits $Q_0$ and $Q_1$, both taking no input qubits
and producing $m$ output qubits.
Let $\rho_0$ and $\rho_1$ be the density matrices corresponding to the
outputs of these circuits.\\[2mm]
{\it Yes:} &
$\delta(\rho_0,\rho_1) \geq 2/3$.\\[2mm]
{\it No:} & 
$\delta(\rho_0,\rho_1) \leq 1/3$.
\end{tabular*}
\end{trivlist}


\begin{theorem}
The quantum state distinguishability problem is \class{QSZK}-complete
with respect to Karp reductions.
\end{theorem}

\noindent
A quantum analogue of a different \class{SZK-complete} problem known
as the {\it entropy difference problem} \cite{GoldreichV99} has
recently been shown to be complete for \class{QSZK} \cite{Ben-AroyaT-S07}.

Kobayashi \cite{Kobayashi08} has recently proved several interesting
properties of quantum {\it computational} zero-knowledge proof systems.

\subsection{Multiple-prover quantum interactive proofs}

Multiple-prover interactive proof systems are variants of interactive
proofs where a verifier interacts with two or more provers that are
not able to communicate with one another during the course of the
interaction.
Classical multiple-prover interactive proofs are extremely powerful: the
class of promise problems having multiple-prover interactive proofs
(denoted \class{MIP}) coincides with \class{NEXP} \cite{BabaiFL91}.
This is true even for two-prover interactive proofs wherein the
verifier exchanges just one round of communication with each prover in
parallel \cite{FeigeL92}.
The key to the power of multiple-prover interactive proofs is the
inability of the provers to communicate during the execution of the
proof system.
Similar to a detective interrogating two suspects in separate rooms in
a police station, the verifier can ask questions of the provers that
require strongly correlated answers, limiting the ability of cheating
provers to convince the verifier of a false statement.

The identification of \class{MIP} with \class{NEXP} assumes a
completely classical description of the provers. 
Quantum information, however, delivers a surprising twist for this
model: even when the verifier is classical, an entangled quantum state
shared between two provers can allow for non-classical correlations
between their answers to the verifier's questions.
This phenomenon is better known as a Bell-inequality violation
\cite{Bell64} in the quantum physics literature.
Indeed, there exist two-prover interactive proof systems that are
sound against classical provers, but can be cheated by entangled
quantum provers \cite{CleveH+04}.

\begin{center}
  \begin{tabular}{p{0.6in}p{5.4in}}
    {\bf $\class{MIP}^{\ast}$} & 
    A promise problem $A = (A_{\yes},A_{\no})$ is in
    $\class{MIP}^{\ast}$ if and only if there exists a multiple-prover
    interactive proof system for $A$ wherein the verifier is classical
    and the provers may share an arbitrary entangled state.
  \end{tabular}
\end{center}

\noindent
One may also consider fully quantum variants of multiple-prover
interactive proofs, which were first studied by Kobayashi and Matsumoto
\cite{KobayashiM03}.

\begin{center}
  \begin{tabular}{p{0.6in}p{5.4in}}
  {\bf $\class{QMIP}$} & 
  A promise problem $A = (A_{\yes},A_{\no})$ is in
  $\class{QMIP}$ if and only if there exists a multiple-prover
  quantum interactive proof system for $A$.
  \end{tabular}
\end{center}

\noindent
Various refinements on these classes have been studied, where
parameters such as the number of provers, number of rounds of
communication, completeness and soundness probabilities, and bounds on
the amount of entanglement shared between provers are taken into
account.

The results proved in both \cite{CleveH+04} and \cite{KobayashiM03} 
support the claim that it is entanglement shared between provers that
is the key issue in multiple-prover quantum interactive proofs.
Both models are equivalent in power to \class{MIP} when provers are
forbidden to share entanglement before the proof system is executed.

At the time of the writing of this article, research into these
complexity classes and general properties of multiple-prover quantum
interactive proofs is highly active and has led to several interesting
results (such as \cite{CleveSUU07,KempeRT07,KempeKMTV07,KempeKMV08},
among others).
Despite this effort, little can be said at this time about the
relationship among the above classes and other known complexity
classes.
For instance, only the trivial lower bounds
$\class{PSPACE}\subseteq\class{MIP}^{\ast}$ and
$\class{QIP}\subseteq\class{QMIP}$, and no good upper bounds, are
known.
It has not even been ruled out that non-computable languages could
have multiple-prover quantum interactive proof systems.
These difficulties seem to stem from two issues:
(i) no bounds are known for the size of entangled states needed for
provers to perform well in an interactive proof, and 
(ii) the possible correlations that can be induced with entanglement
are not well-understood.

One highly restricted variant of $\class{MIP}^{\ast}$ that has been
studied, along with its unentangled counterpart, is as follows.

\begin{center}
\begin{tabular}{p{0.6in}p{5.4in}}
  {\bf $\oplus\class{MIP}^{\ast}$} & 
  A promise problem $A = (A_{\yes},A_{\no})$ is in
  $\oplus\class{MIP}^{\ast}$ if and only if there exists a one-round
  two-prover interactive proof system for $A$ wherein the provers each
  send a single bit to the verifier, and the verifier's decision to
  accept or reject is determined by the questions asked along with the
  XOR of these bits.
  The verifier is classical and the provers are quantum and share an
  arbitrary entangled state.\\[3mm]
  {\bf $\oplus\class{MIP}$} & 
    This class is similar to $\oplus\class{MIP}^{\ast}$, except that
  the provers may not share entanglement (and therefore can be assumed
  to be classical without loss of generality).
\end{tabular}
\end{center}

\noindent
It holds that $\oplus\class{MIP} = \class{NEXP}$ for some choices of
completeness and soundness probabilities \cite{Hastad01,BellareG+98}.
On the other hand, it has been proved \cite{Wehner06} that
$\oplus\class{MIP}^{\ast}\subseteq \class{QIP}(2)$ and therefore
$\oplus\class{MIP}^{\ast}\subseteq\class{EXP}$.

\subsection{Other variants of  quantum interactive proofs}

Other variants of quantum interactive proof systems have been
studied, including {\it public-coin} quantum interactive proofs
and quantum interactive proofs with {\it competing provers}.

\subsubsection*{Public-coin quantum interactive proofs}

Public-coin interactive proof systems are a variant of interactive
proofs wherein the verifier performs no computations until all
messages with the prover have been exchanged.
In place of the verifier's messages are sequences of random bits,
visible to both the prover and verifier.
Such interactive proof systems are typically called
{\it Arthur--Merlin games} \cite{Babai85, BabaiM88}, and the verifier
and prover are called Arthur and Merlin, respectively, in this setting.

Quantum variants of such proof systems and their corresponding classes
are easily defined.
For instance, \class{QAM} is defined as follows \cite{MarriottW05}.
\vspace{3mm}

\noindent
\begin{tabular}{p{0.6in}p{5.5in}}
  {\bf $\class{QAM}$} & 
  A promise problem $A = (A_{\yes},A_{\no})$ is in
  $\class{QAM}$ if and only if it has a quantum interactive
  proof system of the following restricted form.
  Arthur uniformly chooses some polynomial-bounded number of classical
  random bits, and sends a copy of these bits to Merlin.
  Merlin responds with a quantum state on a polynomial-bounded number
  of qubits.
  Arthur then performs a polynomial-time quantum computation on the
  input, the random bits, and the state sent by Merlin to determine
  acceptance or rejection.
\end{tabular}

\vspace{3mm}

\noindent
It is clear that $\class{QAM} \subseteq \class{QIP}(2)$, but unlike
the classical case \cite{GoldwasserS89} equality is not known in the
quantum setting.
It is straightforward to prove the upper bound
$\class{QAM} \subseteq \class{PSPACE}$, while containment
$\class{QIP}(2)\subseteq\class{PSPACE}$ is not known.

The class $\class{QMAM}$ may be defined through a similar analogy with
classical Arthur--Merlin games.
Here, Merlin sends the first message, Arthur responds with a selection
of random bits, and then Merlin sends a second message.

\begin{center}
\begin{tabular}{p{0.6in}p{5.4in}}
  {\bf $\class{QMAM}$} & 
  A promise problem $A = (A_{\yes},A_{\no})$ is in
  $\class{QMAM}$ if and only if it has a quantum interactive
  proof system of the following restricted form.
  Merlin sends a polynomial-bounded number of qubits to Arthur.
  Without performing any computations, Arthur uniformly chooses some
  polynomial-bounded number of classical random bits, and sends a copy
  of these bits to Merlin.
  Merlin responds with a second collection of qubits.
  Arthur then performs a polynomial-time quantum computation on the input,
  the random bits, and the quantum information sent by Merlin in order
  to determine acceptance or rejection.
\end{tabular}
\end{center}

\noindent
Even when Arthur is restricted to a single random bit, this class has
the full power of \class{QIP} \cite{MarriottW05}.

\begin{theorem}
$\class{QMAM} = \class{QIP}$.
\end{theorem}

\subsubsection*{Quantum interactive proofs with competing provers}

Another variant of interactive proof systems that has been studied is
one where a verifier interacts with two {\it competing} provers,
sometimes called the {\it yes-prover} and the {\it no-prover}.
Unlike ordinary two-prover interactive proof systems, it is now
assumed that the provers have conflicting goals: the yes-prover wants
to convince the verifier that a given input string is a yes-instance
of the problem being considered, while the no-prover wants to convince
the verifier that the input is a no-instance.
The verifier is sometimes called the {\it referee} in this setting,
given that interactive proof systems of this form are naturally
modeled as competitive games between the two players.
Two complexity classes based on interactive proofs with competing
provers are the following.

\begin{center}
\begin{tabular}{p{0.6in}p{5.4in}}
  {\bf $\class{RG}$} & 
  A promise problem $A = (A_{\yes},A_{\no})$ is in
  $\class{RG}$ (short for {\it refereed games}) if and only if it has
  a classical interactive proof system with two competing provers.
  The completeness and soundness conditions for such a proof system
  are replaced by the following conditions:
  \begin{enumerate}
    \item
      For every $x\in A_{\yes}$, there exists a yes-prover $P_{\yes}$
      that convinces the referee to {\it accept} with probability at
      least 2/3, regardless of the strategy employed by the no-prover
      $P_{\no}$.

    \item
      For every $x\in A_{\no}$, there exists a no-prover $P_{\no}$
      that convinces the referee to {\it reject} with probability at
      least 2/3, regardless of the strategy employed by the yes-prover
      $P_{\yes}$. 
  \end{enumerate}\\[2mm]
  {\bf $\class{QRG}$} & 
  A promise problem $A = (A_{\yes},A_{\no})$ is in
  $\class{QRG}$ ({\it quantum refereed games}) if and only if it has a
  quantum interactive proof system with two competing provers.
  The completeness and soundness conditions for such a proof system
  are analogous to $\class{RG}$.
\end{tabular}
\end{center}

\noindent
Classical refereed games have the computational power of
deterministic exponential time \cite{FeigeK97}, and the same is true
in the quantum setting \cite{GutoskiW07}:
$\class{RG} = \class{QRG} = \class{EXP}$.
The containment $\class{EXP}\subseteq\class{RG}$ represents an
application of the arithmetization technique, while 
$\class{QRG} \subseteq \class{EXP}$ exemplifies the power
of semidefinite programming.

\section{Other selected notions in quantum complexity}

In this section of this article, a few other topics in quantum
computational complexity theory are surveyed that do not fall under
the headings of the previous sections.
While incomplete, this selection should provide the reader with some
sense for the topics that have been studied in quantum complexity.

\subsection{Quantum advice}

{\it Quantum advice} is a formal abstraction that addresses this
question:
how powerful is quantum software?
More precisely, let us suppose that some polynomial-time generated
family of quantum circuits $Q=\{Q_n\,:\,n\in\natural\}$ is given. 
Rather than assuming that each circuit $Q_n$ takes $n$ input qubits as
for the class \class{BQP}, however, it is now assumed that $Q_n$ takes
$n+p(n)$ qubits for $p$ some polynomial-bounded function: 
in addition to a given input $x\in\Sigma^n$, the circuit $Q_n$ will
take an {\it advice state} $\rho_n$ on $p(n)$ qubits.
This advice state may be viewed as pre-loaded quantum software for a
quantum computer.
The advice state may depend on the input length $n$, but not on
the particular input $x\in\Sigma^n$.
Similar to a quantum proof, the difficulty of preparing this state is
ignored; but unlike a quantum proof the advice state is completely
trusted.
The quantum complexity class $\class{BQP/qpoly}$ is now defined to be
the class of promise problems that are solved by polynomial-time
quantum algorithms with quantum advice in the natural way.
The first published paper on quantum advice was \cite{NishimuraY04},
while the definitions and most of the results discussed in this
section are due to Aaronson \cite{Aaronson05}.

\begin{center}
\begin{tabular}{p{0.8in}p{5.3in}}
  {\bf $\class{BQP/qpoly}$} & 
  A promise problem $A = (A_{\yes},A_{\no})$ is in $\class{BQP/qpoly}$
  if and only if there exists a polynomial-bounded function $p$, a
  collection of quantum states $\{\rho_n\,:\,n\in\natural\}$ where
  each $\rho_n$ is a $p(n)$-qubit state, and a polynomial-time
  generated family of quantum circuits $Q = \{Q_n\,:\,n\in\natural\}$
  with the following properties.
  For all $x\in A_{\yes}$, $Q$ accepts $(x,\rho_{\abs{x}})$ with
  probability at least $2/3$; and for all $x\in A_{\no}$, $Q$ accepts
  $(x,\rho_{\abs{x}})$ with probability at most $1/3$. 
\end{tabular}
\end{center}

\noindent
Similar to \class{BQP} without advice, it is straightforward to shown
that the constants 2/3 and 1/3 in this definition can be replaced by a
wide range of functions $a,b:\natural\rightarrow[0,1]$.

As the notation suggests, \class{BQP/qpoly} is a quantum analogue of
the class \class{P/poly}.
This analogy may be used as the basis for several relevant points
about \class{BQP/qpoly}, quantum advice, and their relationship to
classical complexity classes and notions.
\begin{enumerate}
\item
  As is well-known, \class{P/poly} may be defined either in a manner
  similar to the above definition for \class{BQP/qpoly}, or more simply
  as the class of promise problems solvable by polynomial-size Boolean
  circuit families with no uniformity restrictions.
  Based on similar ideas, the class \class{BQP/qpoly} does not change
  if the circuit family $\{Q_n\,:\,n\in\natural\}$ is taken to be an
  arbitrary polynomial-size circuit family without uniformity
  constraints.
  
\item
  There is a good argument to be made that quantum advice is better
  viewed as a quantum analogue of {\it randomized advice} rather than
  {\it deterministic advice}.
  That is, \class{BQP/qpoly} can equally well be viewed as a quantum
  analogue of the (suitably defined) complexity class \class{BPP/rpoly}.
  It happens to be the case, however, that
  $\class{BPP/rpoly}=\class{P/poly}$.
  (The situation is rather different for 
  {\it logarithmic-length advice}, where randomized advice is strictly
  more powerful than ordinary deterministic advice.)

\item
  Any combination of quantum, randomized, or deterministic advice with
  quantum, randomized, or deterministic circuits can be
  considered.
  This leads to classes such as \class{BQP/rpoly}, \class{BQP/poly},
  \class{P/qpoly}, \class{P/rpoly}, and so on.
  (The only reasonable interpretation of \class{BPP/qpoly} and
  \class{P/qpoly} is that classical circuits effectively measure
  quantum states in the standard basis the instant that they touch
  them.)

  At most three distinct classes among these possibilities arise:
  \class{BQP/qpoly}, \class{BQP/poly}, and \class{P/poly}.
  This is because $\class{BQP/rpoly} = \class{BQP/poly}$ and
  \[
  \class{BPP/qpoly} = \class{BPP/rpoly} = \class{BPP/poly} =
  \class{P/qpoly} = \class{P/rpoly} =  \class{P/poly}.
  \]
  The principle behind these equalities is that nonuniformity
  is stronger than randomness \cite{Adleman78}.
  
\end{enumerate}

The following theorem places an upper-bound on the power of
polynomial-time quantum algorithms with quantum advice \cite{Aaronson05}.

\begin{theorem}
$\class{BQP/qpoly} \subseteq \class{PP/poly}$.
\end{theorem}

\noindent
Although in all likelihood the class \class{PP/poly} is enormous,
containing many interesting problems that one can have little hope of
being able to solve in practice, the upper-bound represented by this
theorem is far from obvious.
The most important thing to notice is that the power of quantum advice
(to a \class{BQP} machine) is simulated by deterministic advice (to a
\class{PP} machine). 
This means that no matter how complex, a polynomial-size quantum
advice state can never encode more information accessible to a
polynomial-time quantum computer than a polynomial-length string can,
albeit to an unbounded error probabilistic machine.

Quantum advice has also been considered for other quantum complexity
classes such as \class{QMA} and \class{QIP}.
For instance, the following bound is known on the power of \class{QMA}
with quantum advice~\cite{Aaronson06}.

\begin{theorem}
$\class{QMA/qpoly} \subseteq \class{PSPACE/poly}$.
\end{theorem}

Generally speaking, the study of both quantum and randomized advice is
reasonably described as a sticky business when unbounded error
machines or interactive protocols are involved.
In these cases, complexity classes defined by both quantum and
randomized advice can be highly non-intuitive and dependent on
specific interpretations of models. 
For example, Raz \cite{Raz05} has shown that both $\class{QIP/qpoly}$
and $\class{IP/rpoly}$ contain all languages, provided that the
advice is not accessible to the prover.
Likewise, Aaronson \cite{Aaronson05} has observed that both
\class{PP/rpoly} and \class{PQP/qpoly} contain all languages.
These results are quite peculiar, given that significantly more
powerful models can become strictly less powerful in the presence of
quantum or randomized advice.
For instance, even a Turing machine running in exponential space
cannot decide all languages with bounded error given randomized
advice.

\subsection{Space-bounded quantum computation}

The quantum complexity classes that have been discussed thus far in
this article are based on the abstraction that efficient quantum
computations are those that can be performed in polynomial time. 
Quantum complexity classes may also be defined by bounds on space
rather than time, but here a different computational model is
required to reasonably compare with classical models of space-bounded
computation.
One simple choice of a suitable model is a hybrid between quantum
circuits and classical Turing machines---and although this model is
different from the variants of quantum Turing machines that were
originally studied in the theory of quantum computing
\cite{Deutsch85,BernsteinV97}, the term {\it quantum Turing machine}
(QTM for short) is nevertheless appropriate.
A different quantum Turing machine model that is suitable for studying
time-space trade-offs has recently been proposed in \cite{MelkebeekW07}.

\begin{figure}[t]
  \begin{center}
    \unitlength=0.9pt
    \begin{picture}(450, 190)(-30,35)

    \node[Nw=50,Nh=50,Nmr=0](Control)(0,200){\makebox(0,0){%
	\begin{tabular}{c}
	  finite\\[-1mm] control
	\end{tabular}
      }}

    \node[Nw=15,Nh=15,Nmr=0](In1)(80,180){}
    \node[Nw=15,Nh=15,Nmr=0](In2)(95,180){}
    \node[Nw=15,Nh=15,Nmr=0](In3)(110,180){}
    \node[Nw=15,Nh=15,Nmr=0](In4)(125,180){}
    \node[Nw=15,Nh=15,Nmr=0](In5)(140,180){}
    \node[Nw=15,Nh=15,Nmr=0](In6)(155,180){}
    \node[Nw=15,Nh=15,Nmr=0](In7)(170,180){}
    \node[Nw=15,Nh=15,Nmr=0](In8)(185,180){}
    \node[Nw=15,Nh=15,Nmr=0](In9)(200,180){}
    \node[Nw=15,Nh=15,Nmr=0](In10)(215,180){}
    \node[Nw=15,Nh=15,Nmr=0](In11)(230,180){}
    \node[Nw=15,Nh=15,Nmr=0](In12)(245,180){}
    \node[Nw=15,Nh=15,Nmr=0](In13)(260,180){}
    \node[Nw=15,Nh=15,Nmr=0](In14)(275,180){}
    \node[Nw=15,Nh=15,Nmr=0](In15)(290,180){}

    \node[Nw=15,Nh=15,Nmr=0](Work1)(80,115){}
    \node[Nw=15,Nh=15,Nmr=0](Work2)(95,115){}
    \node[Nw=15,Nh=15,Nmr=0](Work3)(110,115){}
    \node[Nw=15,Nh=15,Nmr=0](Work4)(125,115){}
    \node[Nw=15,Nh=15,Nmr=0](Work5)(140,115){}
    \node[Nw=15,Nh=15,Nmr=0](Work6)(155,115){}
    \node[Nw=15,Nh=15,Nmr=0](Work7)(170,115){}
    \node[Nw=15,Nh=15,Nmr=0](Work8)(185,115){}
    \node[Nw=15,Nh=15,Nmr=0](Work9)(200,115){}
    \node[Nw=15,Nh=15,Nmr=0](Work10)(215,115){}
    \node[Nframe=n,Nw=3,Nh=3,fillcolor=Black](Dot)(230,115){}
    \node[Nframe=n,Nw=3,Nh=3,fillcolor=Black](Dot)(240,115){}
    \node[Nframe=n,Nw=3,Nh=3,fillcolor=Black](Dot)(250,115){}
    
    \node[Nw=15,Nh=15,Nmr=0](Qubit1)(80,50){}
    \node[Nw=15,Nh=15,Nmr=0](Qubit2)(95,50){}
    \node[Nw=15,Nh=15,Nmr=0](Qubit3)(110,50){}
    \node[Nw=15,Nh=15,Nmr=0](Qubit4)(125,50){}
    \node[Nw=15,Nh=15,Nmr=0](Qubit5)(140,50){}
    \node[Nw=15,Nh=15,Nmr=0](Qubit6)(155,50){}
    \node[Nw=15,Nh=15,Nmr=0](Qubit7)(170,50){}
    \node[Nw=15,Nh=15,Nmr=0](Qubit8)(185,50){}
    \node[Nw=15,Nh=15,Nmr=0](Qubit9)(200,50){}
    \node[Nw=15,Nh=15,Nmr=0](Qubit10)(215,50){}
    \node[Nframe=n,Nw=3,Nh=3,fillcolor=Black](Dot)(230,50){}
    \node[Nframe=n,Nw=3,Nh=3,fillcolor=Black](Dot)(240,50){}
    \node[Nframe=n,Nw=3,Nh=3,fillcolor=Black](Dot)(250,50){}

    \drawline[arcradius=20](25,210)(200,210)(200,190)
    \drawline[arcradius=20](25,200)(50,200)(60,135)(170,135)(170,125)
    \drawline[arcradius=20](25,190)(40,190)(60,80)(170,80)(170,60)
    \drawline[arcradius=20](90,80)(125,80)(125,60)
    \drawline[arcradius=20](90,80)(140,80)(140,60)

    \put(310,180){\makebox(0,0)[l]{input tape (read-only)}}
    \put(280,115){\makebox(0,0)[l]{classical work tape}}
    \put(280,50){\makebox(0,0)[l]{qubit tape}}

    \end{picture}
  \end{center}

  \caption{A quantum Turing machine.
    This variant of quantum Turing machine is classical with the
    exception of a quantum work tape, each square of which contains a
    qubit.
    Quantum operations and measurements can be performed on the qubits
    scanned by the quantum tape heads.}
  \label{fig:QTM}
\end{figure}
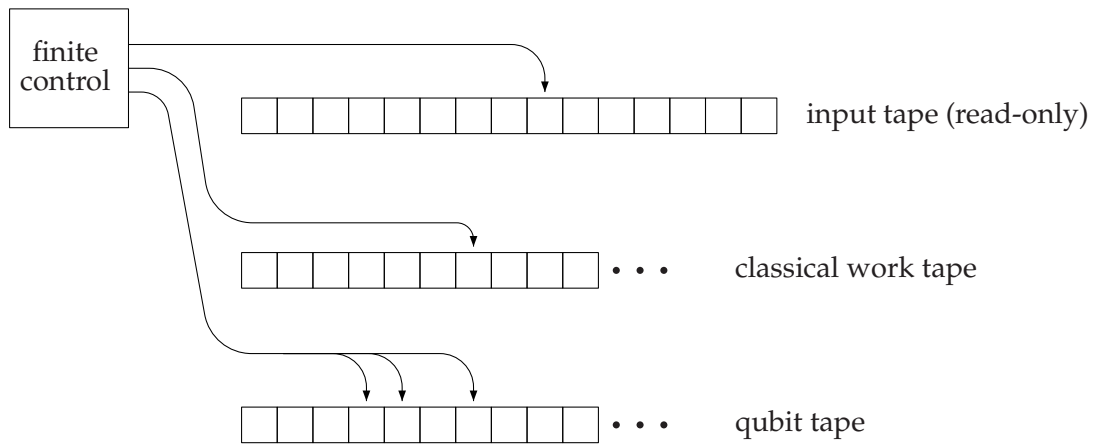

Figure~\ref{fig:QTM} illustrates a quantum Turing machine.
A quantum Turing machine has a read-only classical input tape, a
classical work tape, and a quantum tape consisting of an infinite
sequence of qubits each initialized to the zero-state.
Three tape heads scan the quantum tape, allowing quantum operations
to be performed on the corresponding qubits.
(It would be sufficient to have just two, but allowing three tape
heads parallels the choice of a universal gate set that allows
three-qubit operations.)
A single step of a QTM's computation may involve ordinary moves by the
classical parts of the machine and quantum operations on the quantum
tape: Toffoli gates, Hadamard gates, phase-shift gates, or
single-qubit measurements in the computational basis.

The running time of a quantum Turing machine is defined as for
ordinary Turing machines.
The space used by such a machine is the number of squares on the
classical work tape plus the number of qubits on the quantum tape that
are ever visited by one of the tape heads.
Similar to the classical case, the input tape does not contribute to
the space used by the machine because it is a read-only tape.

The following complexity classes are examples of classes that can be
defined using this model.

\begin{center}
  \begin{tabular}{p{0.9in}p{5.1in}}
    {\bf \class{BQL}} & 
    A promise problem $A = (A_{\yes},A_{\no})$ is in \class{BQL}
    (bounded-error quantum logarithmic space) 
    if and only if there exists a quantum Turing machine $M$ running in
    polynomial time and logarithmic space that accepts every string
    $x\in A_{\yes}$ with probability at least $2/3$ and accepts every
    string $x\in A_{\no}$ with probability at most $1/3$.\\[2mm]
  \end{tabular}\\
  \begin{tabular}{p{0.9in}p{5.1in}}
    {\bf \class{PQL}} & 
    A promise problem $A = (A_{\yes},A_{\no})$ is in \class{PQL}
    (unbounded-error quantum logarithmic space) 
    if and only if there exists a quantum Turing machine $M$ running in
    polynomial time and logarithmic space that accepts every string
    $x\in A_{\yes}$ with probability strictly greater than $1/2$ and
    accepts every string $x\in A_{\no}$ with probability at most
    $1/2$.\\[2mm] 
  \end{tabular}
  \begin{tabular}{p{0.9in}p{5.1in}}
    {\bf \class{BQPSPACE}} & 
    A promise problem $A = (A_{\yes},A_{\no})$ is in \class{BQPSPACE}
    (bounded-error quantum polynomial space) 
    if and only if there exists a quantum Turing machine $M$ running in
    polynomial space that accepts every string $x\in A_{\yes}$ with
    probability at least $2/3$ and accepts every string
    $x\in A_{\no}$ with probability at most $1/3$.\\[2mm] 
  \end{tabular}
  \begin{tabular}{p{0.9in}p{5.1in}}
    {\bf \class{PQPSPACE}} & 
    A promise problem $A = (A_{\yes},A_{\no})$ is in \class{PQPSPACE} 
    (unbounded-error quantum polynomial space) 
    if and only if there exists a quantum Turing machine $M$ running in
    polynomial space that accepts every string $x\in A_{\yes}$ with
    probability strictly greater than $1/2$ and accepts every string
    $x\in A_{\no}$ with probability at most $1/2$.\\[2mm] 
  \end{tabular}
\end{center}

Unlike polynomial-time computations, it is known that quantum
information does not give a significant increase in computational
power in the space-bounded case \cite{Watrous99,Watrous03}.

\begin{theorem}
  The following relationships hold.
  \begin{enumerate}
    \item
      $\class{BQL} \subseteq \class{PQL} = \class{PL}$.
    \item
      $\class{BQPSPACE} = \class{PQPSPACE} = \class{PSPACE}$.
  \end{enumerate}
\end{theorem}

\noindent
The key relationship in the above theorem, from the perspective of
quantum complexity, is $\class{PQL} \subseteq \class{PL}$, which can
be shown using space-bounded counting complexity.
In particular, the proof relies on a theory of \class{GapL} functions
\cite{AllenderO96} that parallels the theory of \class{GapP}
functions, and allows for a variety of matrix computations to be
performed in \class{PL}.

The above theorem, together with the containment
$\class{PL}\subseteq\class{NC}$ \cite{BorodinC+83}, implies
that both $\class{BQL}$ and $\class{PQL}$ are contained in
$\class{NC}$.
An interpretation of this fact is that logarithmic-space quantum
computations can be very efficiently simulated in parallel.

\subsection{Bounded-depth quantum circuits}

The {\it depth} of a classical or quantum circuit is the maximum number
of gates encountered on any path from an input bit or qubit to an
output bit or qubit in the circuit.
One may reasonably think of circuit depth as the parallel running
time, or the number of time units needed to apply a circuit when
operations may be parallelized in any way that respects the topology
of the circuit.

The following complexity class, first defined by
Moore and Nilsson \cite{MooreN02}, represent bounded-error quantum
variants of the class $\class{NC}$.

\begin{center}
\begin{tabular}{p{0.6in}p{5.4in}}
  {\bf $\class{QNC}$} & 
  A promise problem $A = (A_{\yes},A_{\no})$ is in $\class{QNC}$
  (bounded-error quantum \class{NC}) if and only if there exists 
  a logarithmic-space generated family $\{Q_n\,:\,n\in\natural\}$
  of poly-logarithmic depth quantum circuits, where each circuit $Q_n$
  takes $n$ input qubits and produces one output qubit, such that
  $\op{Pr}[\text{$Q$ accepts $x$}]\geq 2/3$ for all $x\in A_{\yes}$, 
  and $\op{Pr}[\text{$Q$ accepts $x$}]\leq 1/3$ for all $x\in A_{\no}$.
\end{tabular}
\end{center}

\noindent
Many other complexity classes based on bounded-depth quantum circuits
have been studied as well.
The survey of Bera, Green, and Homer \cite{BeraGH07} discusses several
examples.

In the classical case there is a very close relationship between
space-bounded and depth-bounded computation
\cite{Borodin77,BorodinC+83}. 
This close relationship is based on two main ideas:
the first is that space-bounded computations can be simulated
efficiently by bounded-depth circuits using parallel algorithms
for matrix computations, and the second is that bounded-depth Boolean
circuits can be efficiently simulated by space-bounded computations
via depth-first traversals of the circuit to be simulated.

For quantum computation this close relationship is not known to exist.
One direction indeed holds, as was discussed in the previous subsection:
space-bounded quantum computations can be efficiently simulated by
depth-bounded circuits.
The other direction, which is an efficient space-bounded simulation of
bounded-depth quantum circuits, is not known to hold and is arguably
quite unlikely.
Informally speaking, bounded-depth quantum circuits are
computationally powerful, whereas space-bounded quantum Turing
machines are not.
Three facts that support this claim are as follows.
\begin{enumerate}
\item
Computing the acceptance probability for even constant-depth quantum
circuits is as hard as computing acceptance probabilities for
arbitrary polynomial-size quantum circuits \cite{FennerG+05}.

\item
Shor's factoring algorithm can be implemented by quantum circuits
having logarithmic-depth, along with classical pre- and post-processing
\cite{CleveW00}. 

\item
The quantum circuit distinguishability problem remains complete for
$\class{QIP}$ when restricted to logarithmic-depth quantum circuits
\cite{Rosgen08}.
\end{enumerate}

\noindent
It is reasonable to conjecture that $\class{QNC}$ is incomparable
with $\class{BPP}$ and properly contained in $\class{BQP}$.

\section{Future directions}

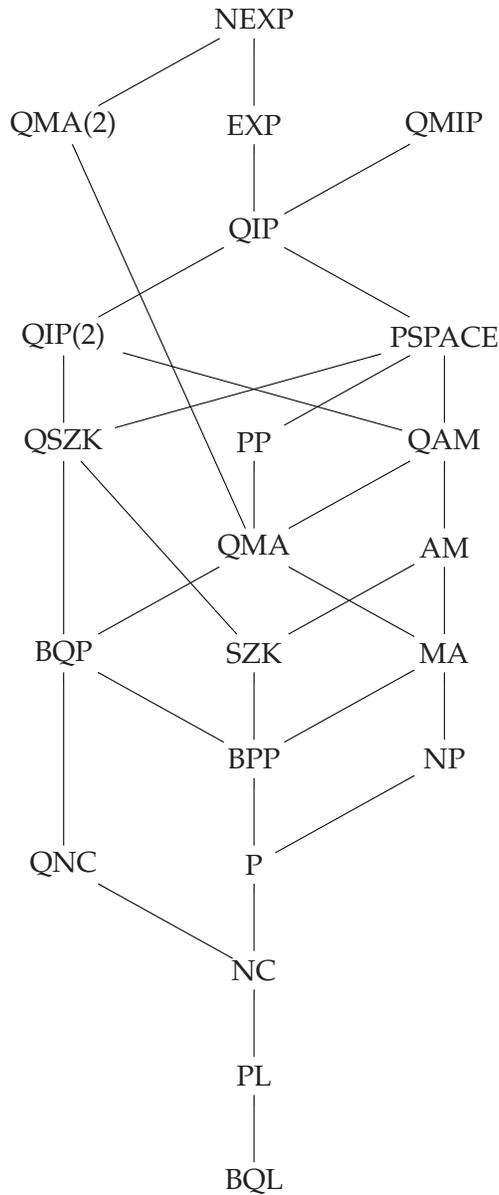
\begin{figure}[!t]
\begin{center}
  \unitlength=1.6pt
  \begin{picture}(140, 290)(-20,-80)
    \gasset{AHLength=0,ilength=-6,Nh=9,Nmr=0,Nframe=n}

    \node[Nw=10](P)(50,0){\class{P}}
    \node[Nw=24](QNC)(5,0){\class{QNC}}
    
    \node(NP)(95,25){\class{NP}}
    \node(PP)(50,100){\class{PP}}
    \node(BPP)(50,25){\class{BPP}}
    \node(MA)(95,50){\class{MA}}
    \node(AM)(95,75){\class{AM}}
    \node[Nw=30](PSPACE)(95,125){\class{PSPACE}}
    \node(EXP)(50,175){\class{EXP}}
    \node[Nw=24](NEXP)(50,200){\class{NEXP}}
    \node[Nw=18](BQP)(5,50){\class{BQP}}
    \node[Nw=20](QMA)(50,75){\class{QMA}}
    \node[Nw=24](QSZK)(5,100){\class{QSZK}}
    \node(QIP)(50,150){\class{QIP}}
    \node[Nw=28](QIP2)(5,125){\class{QIP}(2)}
    \node[Nw=20](QAM)(95,100){\class{QAM}}
    \node[Nw=24](QMIP)(95,175){\class{QMIP}}
    \node[Nw=18](SZK)(50,50){\class{SZK}}
    \node[Nw=10](NC)(50,-25){\class{NC}}
    \node[Nw=10](PL)(50,-50){\class{PL}}
    \node[Nw=10](BQL)(50,-75){\class{BQL}}
    \node[Nw=24](QMMA)(5,175){\class{QMA(2)}}

    \drawedge(QMA,QMMA){}
    \drawedge(QMMA,NEXP){}    
    \drawedge(BQL,PL){}
    \drawedge(PL,NC){}
    \drawedge(NC,P){}
    \drawedge(NC,QNC){}
    \drawedge(EXP,NEXP){}
    \drawedge(QIP,EXP){}
    \drawedge(QIP2,QIP){}
    \drawedge(PSPACE,QIP){}    
    \drawedge(QSZK,QIP2){}    
    \drawedge(QSZK,PSPACE){}    
    \drawedge(QIP,QMIP){}    
    \drawedge(QAM,QIP2){}
    \drawedge(P,BPP){}
    \drawedge(P,NP){}
    \drawedge(BPP,MA){}
    \drawedge(NP,MA){}
    \drawedge(MA,AM){}
    \drawedge(MA,QMA){}
    \drawedge(QMA,PP){}
    \drawedge(QMA,QAM){}
    \drawedge(BQP,QMA){}
    \drawedge(BQP,QSZK){}
    \drawedge(AM,QAM){}
    \drawedge(QAM,PSPACE){}
    \drawedge(PP,PSPACE){}
    \drawedge(BPP,BQP){}
    \drawedge(QNC,BQP){}
    \drawedge(BPP,SZK){}
    \drawedge(SZK,QSZK){}
    \drawedge(SZK,AM){}
  \end{picture}
\end{center}

\caption{A diagram of inclusions among most of the complexity classes
  discussed in this article.}
\label{fig:classes}
\end{figure}

There are many future directions for research in quantum
computational complexity theory.
The following list suggests just a few of many possibilities.

\begin{enumerate}
\item
The power of multiple-prover quantum interactive proofs, for both quantum
and classical verifiers, is very poorly understood.
In particular:
(i) no interesting upper-bounds are known for either
$\class{MIP}^{\ast}$ or \class{QMIP}, and 
(ii) neither the containment
$\class{NEXP} \subseteq \class{MIP}^{\ast}$ nor
$\class{NEXP} \subseteq \class{QMIP}$
is known to hold.

\item
The containment $\class{NP}\subseteq\class{BQP}$ would be a very
powerful incentive to build a quantum computer, to say the least.
While there is little reason to hope that this containment holds,
there is at the same time little evidence against it aside from the
fact that it fails relative to an oracle \cite{BennettBBV97}.
A better understanding of the relationship between \class{BQP} and
\class{NP}, including possible consequences of one being contained in
the other, is an interesting direction for further research in quantum
complexity.

\item
Along similar lines to the previous item, an understanding of the
relationship between \class{BQP} and the polynomial-time hierarchy has
remained elusive. 
It is not known whether \class{BQP} is contained in the
polynomial-time hierarchy, whether there is an oracle relative to
which this is false, or even whether there is an oracle relative to
which \class{BQP} is not contained in \class{AM}.

\item
Interest in complexity classes is ultimately derived from the
problems that they contain.
An important future direction in quantum complexity theory is to prove
the inclusion of interesting computational problems in the quantum
complexity classes for which this is possible.
Of the many problems that have been considered, one of the most
perplexing from the perspective of quantum complexity is the graph
isomorphism problem.
It is a long-standing open problem whether the graph isomorphism
problem is in \class{BQP}.
A seemingly easier task than showing the inclusion of this problem in
\class{BQP} is proving that its complement is contained in
\class{QMA}.
In other words, can Merlin prepare a quantum state that convinces
Arthur that two graphs are not isomorphic?

\end{enumerate}


\bibliographystyle{plain}

\end{document}